\documentclass[]{aa}
\usepackage{epsfig}
\begin{document}
\renewcommand{\topfraction}{1.}
\renewcommand{\bottomfraction}{1.}
\renewcommand{\textfraction}{0.}
\title{Circumstellar masers in the Magellanic Clouds}
\author{Jacco Th. van Loon\inst{1}, Albert A. Zijlstra\inst{2}, Valent\'{\i}n
        Bujarrabal\inst{3}, Lars-{\AA}ke Nyman\inst{4,5}}
\institute{Institute of Astronomy, Madingley Road, Cambridge CB3 0HA, United
           Kingdom
      \and UMIST, P.O.Box 88, Manchester M60 1QD, United Kingdom
      \and Observatorio Astron\'{o}mico Nacional, Campus Universitario,
           Apartado 1143, E-28800 Alcal\'{a} de Henares, Spain
      \and European Southern Observatory, Casilla 19001, Santiago 19, Chile
      \and Onsala Space Observatory, S-439 92 Onsala, Sweden}
\offprints{Jacco van Loon, \email{jacco@ast.cam.ac.uk}}
\date{Received date; accepted date}
\titlerunning{Circumstellar masers in the Magellanic Clouds}
\authorrunning{van Loon et al.}
\abstract{Results are presented of a search for 22 GHz H$_2$O
$6_{16}\rightarrow5_{23}$, 43 GHz SiO$_{\rm v=1}(J=1\rightarrow0)$, 86 GHz
SiO$_{\rm v=1}(J=2\rightarrow1)$ and 129 GHz SiO$_{\rm v=1}(J=3\rightarrow2)$
maser emission from bright IRAS point sources in the Small and Large
Magellanic Clouds --- mostly circumstellar envelopes around obscured red
supergiants and Asymptotic Giant Branch stars (OH/IR stars). The aim of this
effort was to test whether the kinematics of the mass loss from these stars
depends on metallicity.\\
H$_2$O maser emission was detected in the red supergiants IRAS04553$-$6825 and
IRAS05280$-$6910, and tentatively in the luminous IR object IRAS05216$-$6753
and the AGB star IRAS05329$-$6708, all in the LMC. SiO$_{\rm
v=1}(J=2\rightarrow1)$ maser emission was detected in IRAS04553$-$6825.\\
The double-peaked H$_2$O maser line profiles of IRAS04553$-$6825 and
IRAS05280$-$6910, in combination with the OH (and SiO) maser line profiles,
yield the acceleration of the outflows from these stars. The outflow velocity
increases between the H$_2$O masing z\^{o}ne near the dust-formation region
and the more distant OH masing z\^{o}ne from $v\sim18$ to 26 km for
IRAS04553$-$6825 and from $v\sim6$ to 17 km s$^{-1}$ for IRAS05280$-$6910.\\
The total sample of LMC targets is analysed in comparison with circumstellar
masers in the Galactic Centre. The photon fluxes of circumstellar masers in
the LMC are found to be very similar to those in the Galactic Centre. The
expansion velocities in the LMC appear to be $\sim20$\% lower than for
similarly bright OH masers in the Galactic Centre, but the data are still
consistent with no difference in expansion velocity. OH/IR stars in the LMC
appear to have slower accelerating envelopes than OH/IR stars in the Galactic
Centre.\\
The masers in the LMC have blue-asymmetric emission profiles. This may be due
to the amplification of stellar and/or free-free radiation, rather than the
amplification of dust emission, and may be more pronounced in low metallicity
envelopes.\\
The SiO maser strength increases with the photometric amplitude at 2.2 $\mu$m
but is independent of the photometric amplitude at 10 $\mu$m. This suggests a
strong connection between shocks in the dust-free SiO masing z\^{o}ne and the
dust formation process. The LMC masers obey the same trend as the Galactic
Centre masers.\\
Appendices describe H$_2$O maser emission from the moderately mass-losing AGB
star R Dor in the Milky Way, optical echelle spectroscopy of IRAS04553$-$6825,
and the properties of circumstellar masers in the Galactic Centre.
}
\maketitle

\section{Introduction}

In the latest stages of their evolution, both massive and intermediate-mass
stars pass through a phase of intense mass loss at rates of $10^{-6}$ to
$10^{-3}$ M$_\odot$ yr$^{-1}$ (van Loon et al.\ 1999b), returning a major
fraction of their initial mass to the interstellar medium (ISM). For stars of
$M_{\rm initial}\ga8$ M$_\odot$ this occurs when they are red supergiants
(RSGs), and for stars of $1{\la}M_{\rm initial}\la8$ M$_\odot$ when they are
Asymptotic Giant Branch (AGB) stars before becoming a Planetary Nebula (PN).
They become enshrouded by their dusty circumstellar envelope (CSE), rendering
them invisible at optical wavelengths. The absorbed radiation is re-emitted by
the dust at longer wavelengths, making them very bright IR objects. This is
also when, in oxygen-rich CSEs, maser emission from OH, H$_2$O and SiO
molecules may be observed. Hence these objects are known as OH/IR stars.

The locations and intensities of masers are determined by molecular abundance,
dust temperature, gas density, and velocity (Goldreich \& Scoville 1976; Lewis
1989). Where H$_2$O is not shielded from interstellar UV radiation it is
dissociated into OH. SiO is depleted into dust grains except close to the
star. Maser strengths scale with the local kinetic energy density or the local
IR radiation field, which both depend on the CSE temperature. Masers are
quenched above a critical density. Finally, the radiation field in the maser
transition is amplified only by molecules that have small projected velocity
differences (within the thermal width of the maser line, i.e.\ ${\la}1$ km
s$^{-1}$).

As a result the CSE has a layered maser structure, corresponding to subsequent
stages in the mass-loss process from the underlying star. Strong pulsations of
the stellar photosphere of these Long Period Variables (LPVs) eject matter in
which dust forms at typically 1 to 10 stellar radii ($R_\star$). Radiation
pressure accelerates the matter to velocities exceeding the escape velocity.
Matter that does not reach the dust formation radius falls back to the star.
At distances of $\ga100 R_\star$, the stellar wind flows with constant
velocity $v_\infty$ into interstellar space. SiO masers probe the inner
dust-free z\^{o}ne, H$_2$O masers probe the acceleration z\^{o}ne, and OH
masers probe the final stellar wind. This has been beautifully confirmed by
interferometric observations of Galactic OH/IR stars (Diamond et al.\ 1984;
Richards et al.\ 1996; Colomer et al.\ 2000).

The physical conditions in the CSE, and the evolutionary stage of the star can
be determined from the presence or absence of different species of masers, and
from their photon fluxes. The kinematic structure of the CSE can be determined
from the maser line profiles. SiO maser radiation is predominantly amplified
tangentially, resulting in a single maser peak within a few km s$^{-1}$ of the
stellar velocity $v_\star$. OH maser radiation is beamed radially, resulting
in a double-peaked line profile, spanning $2\times v_\infty$. H$_2$O masers
are single-peaked in Mira variables but double-peaked in OH/IR stars (Takaba
et al.\ 1994). Hence, in OH/IR stars H$_2$O masers yield the expansion
velocity at the base of the acceleration z\^{o}ne in the CSE.

%
%
\begin{table*}
\caption[]{Summary of observations: observatory, dates, transition, rest
frequency, centre velocity, channel velocity width, number of channels,
bandwidth, effective telescope diameter, beam Full-Width-Half-Maximum, typical
system temperature and conversion factor. See text for more SEST
observations.}
\begin{tabular}{lrlllllrllcr}
\hline\hline
site                             &
dates                            &
transition                       &
$\nu_0$                          &
$v_{\rm hel, c}$                 &
${\Delta}v$                      &
$N$                              &
band                             &
$D_{\rm eff}$                    &
beam                             &
$T_{\rm sys}$                    &
cal.                             \\
                                 &
d/m/y                            &
                                 &
GHz                              &
km/s                             &
km/s                             &
                                 &
MHz                              &
m                                &
$^\prime$                        &
K                                &
Jy/K                             \\
\hline
Parkes                           &
19--20/8/97                      &
H$_2$O $6_{16}\rightarrow5_{23}$ &
22.23507985                      &
340                              &
0.84                             &
1024                             &
64                               &
45                               &
1.3                              &
110                              &
6                                \\
Parkes                           &
5--13/4/00                       &
H$_2$O $6_{16}\rightarrow5_{23}$ &
22.23507985                      &
200                              &
0.42                             &
2049                             &
64                               &
45                               &
1.3                              &
140--180                         &
6                                \\
Mopra                            &
18--29/1/99                      &
H$_2$O $6_{16}\rightarrow5_{23}$ &
22.23507985                      &
270                              &
0.84                             &
1024                             &
64                               &
22                               &
2.7                              &
115                              &
20                               \\
Parkes                           &
16--25/8/95                      &
SiO $1\rightarrow0$ ($v=1$)      &
43.122080                        &
249.8                            &
0.217\rlap{3}                    &
1024                             &
32                               &
17                               &
1.6                              &
80                               &
39                               \\
Mopra                            &
\llap{2}9/8--6/9/95              &
SiO $2\rightarrow1$ ($v=1$)      &
86.243442                        &
250                              &
0.217\rlap{3}                    &
1024                             &
64                               &
16                               &
1.0                              &
130                              &
45                               \\
SEST                             &
25--31/5/95                      &
SiO $2\rightarrow1$ ($v=1$)      &
86.243442                        &
290                              &
0.15                             &
2000                             &
86                               &
15                               &
0.95                             &
110--150                         &
25                               \\
SEST                             &
29--31/5/95                      &
SiO $2\rightarrow1$ ($v=1$)      &
86.243442                        &
290                              &
2.4                              &
1600                             &
1086                             &
15                               &
0.95                             &
110--150                         &
25                               \\
SEST                             &
25--28/5/95                      &
SiO $3\rightarrow2$ ($v=1$)      &
\llap{1}29.363368                &
290                              &
1.6                              &
1440                             &
995                              &
15                               &
0.67                             &
190--250                         &
29                               \\
\hline
\end{tabular}
\end{table*}

A significant population of confirmed and suspected OH/IR stars in the Small
and Large Magellanic Clouds (SMC \& LMC) has now been identified and studied
(Wood et al.\ 1992; Zijlstra et al.\ 1996; Loup et al.\ 1997; van Loon et al.\
1997, 1998a, 1999a,b). The metallicities of the intermediate-age populations
in the SMC and LMC are $\sim7$ and $\sim3\times$ lower than solar, and hence
the dependence of the mass loss on metallicity may be investigated (van Loon
2000). A study of Magellanic circumstellar masers may shed light on the
metallicity dependence of the envelope kinematics. Here the final results are
presented of surveys for H$_2$O and SiO masers in the Magellanic Clouds (van
Loon et al.\ 1996, 1998b).

\section{Radio observations}

A summary of the observations can be found in Table 1. No attempts were made
to better determine the positions of (tentatively) detected maser sources.

\subsection{H$_2$O maser emission at 22 GHz with Parkes}

The 64 m radio telescope at Parkes, Australia, was used with the 1.3 cm
receiver plus autocorrelator to observe 22 GHz H$_2$O masers. Using the Dual
Circular feed spectra were obtained simultaneously in left and right circular
polarization. No difference was found between them, and they were averaged.
The current K-band facility at Parkes is not as powerful as the beam-switching
set-up used in the early 1980's that yielded $T_{\rm sys}\sim90$ K (Whiteoak
\& Gardner 1986). Weather conditions were generally fair.

In the 1997 run only the LMC targets IRAS04553$-$6825 and IRAS05329$-$6708
were observed, for six and three hours on-source integration, respectively.
The nearby sky was measured every two minutes, resulting in very flat
baselines that required only a shallow second-order polynomial to be
subtracted. The 22 GHz discovery spectrum of IRAS04553$-$6825 was already
presented in van Loon et al.\ (1998b). In the 2000 run, on-source integration
times were a few hours per target. The sky was measured every two minutes ---
most of the time at $5^\prime$ N of the source. Pointing, focus and
calibration were checked regularly by observing the nearby bright maser source
R Dor (see Appendix A) at different zenith distances. The pointing accuracy
was found to be $\sim10^{\prime\prime}$ and flux losses due to pointing errors
are less than $\sim10\%$.

\subsection{H$_2$O maser emission at 22 GHz with Mopra}

The 22 m radio telescope at Mopra, Australia, was used with the 1.3 cm
receiver plus autocorrelator to observe 22 GHz H$_2$O masers. The opacity
correction was usually between 1.3 and 1.6. The conversion factor agrees with
the observed noise and flux density for R Dor (48 Jy).

The background was measured every four minutes at both sides of the target.
The baseline was estimated for each individual spectrum by a running average
(within 1-$\sigma$ around the median) over 55 channels. After subtracting the
baselines, the flux density was obtained by averaging (within 3-$\sigma$
around the median) over the time series of spectra. This procedure yielded
flat baselines and effectively rejected bad spectra. Flux conservation was
tested on the well-detected H$_2$O maser spectra of IRAS04553$-$6825.

\subsection{SiO maser emission at 43 GHz with Parkes}

The 64 m radio telescope at Parkes, Australia, was used with the 0.7 cm single
polarization receiver plus correlator to observe 43 GHz SiO masers. The
background was measured by alternation between two targets every 150 seconds.
Reasonably flat baselines were obtained by subtracting a fifth order
polynomial plus a sine with a period of 200 channels. The useful heliocentric
velocity range runs from 167 to 322 km s$^{-1}$.

\subsection{SiO maser emission at 86 GHz with Mopra}

The 22 m radio telescope at Mopra, Australia, was used with the 3 mm SIS
receiver plus correlator to observe 86 GHz SiO masers. The spectra were
obtained simultaneously in two orthogonal polarizations that were then
averaged. The pointing was checked every few hours on R Dor, and was accurate
to $<15^{\prime\prime}$ rms. The background was measured by alternation
between two of these targets every few minutes. The baselines were very flat.

\subsection{SiO maser emission at 86 \& 129 GHz with SEST}

The 15 m Swedish-ESO Sub-mm Telescope (SEST) at the European Southern
Observatory (ESO) at La Silla, Chile, was used with the 3 mm receiver plus the
acousto-optical High Resolution Spectrograph (HRS) to observe 86 GHz SiO
masers. This configuration was used simultaneously with the Low Resolution
Spectrograph (LRS) either tuned at the same frequency or coupled to the 2 mm
receiver to observe 129 GHz SiO masers.

The internal absolute flux calibration is accurate to $\sim20$\%. The
observations were done in Double Beam Switch Mode, with a beam throw of
$\sim11.5^{\prime}$ in azimuth. The pointing was checked every few hours on R
Dor, and was accurate to $\sim3^{\prime\prime}$ rms. The atmospheric
conditions were very good: a relative humidity of typically 15 to 30\%, an
outside air temperature of $\sim15$ $^\circ{\rm C}$, and little or no cirrus.
The baselines were very flat, requiring only a zeroth order polynomial to be
subtracted from each spectrum.

IRAS04553$-$6825, after its 86 GHz SiO maser emission had been detected (van
Loon et al.\ 1996), was re-observed on 7 occasions during the period from July
1996 to January 1998, with the SEST and the HRS and LRS at 86 GHz using the
same observing strategy as described above. The total on-source integration
time at 86 GHz was 65 and 51 hr with the HRS and LRS, respectively. During
this campaign IRAS05329$-$6708 and IRAS05280$-$6910 were re-observed in
January 1998.

\section{Results from the maser searches}

%
%
\begin{table}
\caption[]{The SMC \& LMC targets, together with their IRAS flux densities (in
Jy), classifications and whether OH 1612 MHz masers are detected (yes), not
detected (no), or not tried ( ).}
\begin{tabular}{lrrlc}
\hline\hline
IRAS-PSC     &
S$_{12}$     &
S$_{25}$     &
type         &
OH           \\
\hline
{\it SMC sources} &
                  &
                  &
                  &
                  \\
00483$-$7347 &
 0.64        &
 0.49        &
AGB          &
             \\
00486$-$7308$^\dagger$ &
   0.41      &
$<$0.60      &
AGB          &
             \\
01074$-$7140 &
 0.36        &
 0.47        &
AGB          &
             \\
\hline
{\it LMC sources} &
                  &
                  &
                  &
                  \\
04407$-$7000 &
 0.81        &
 0.70        &
AGB          &
yes          \\
04491$-$6915 &
 0.53        &
 2.17        &
H {\sc ii}   &
             \\
04498$-$6842 &
 1.31        &
 1.05        &
AGB          &
             \\
04514$-$6931 &
 0.36        &
 3.52        &
H {\sc ii} ? &
             \\
04530$-$6916 &
 2.12        &
 5.10        &
RSG ?        &
no           \\
04545$-$7000 &
 0.52        &
 0.84        &
AGB          &
yes          \\
04546$-$6915 &
 1.17        &
 9.67        &
H {\sc ii}   &
             \\
04553$-$6933 &
 0.53        &
 0.48        &
RSG          &
             \\
04553$-$6825 &
 9.15        &
14.36        &
RSG          &
yes          \\
04571$-$6627 &
 0.42        &
 3.03        &
H {\sc ii}   &
             \\
04581$-$7013 &
 0.41        &
 0.39        &
RSG          &
             \\
05198$-$6941 &
 2.63        &
 7.15        &
H {\sc ii}   &
             \\
05216$-$6753 &
 4.10        &
14.56        &
RSG ?        &
no           \\
05280$-$6910 &
 4.16        &
24.18        &
RSG ?        &
yes          \\
05298$-$6957 &
 0.85        &
 1.38        &
AGB          &
yes          \\
05325$-$6743 &
 1.20        &
 6.47        &
H {\sc ii}   &
             \\
05329$-$6708 &
 0.98        &
 1.48        &
AGB          &
yes          \\
05346$-$6949 &
 7.78        &
20.75        &
RSG          &
no           \\
05402$-$6956 &
 0.82        &
 1.11        &
AGB          &
yes          \\
\hline
\end{tabular}\\
$^\dagger$\footnotesize{IRAS00486$-$7308 is from the IRAS-FSC with flux
densities from Groenewegen \& Blommaert (1998).}
\end{table}

The targets comprise the known Magellanic OH/IR stars, plus other IR objects
in the SMC and LMC: mainly dust-enshrouded AGB stars, RSGs or H {\sc ii}
emission objects (LHA numbers from Henize 1956) often related to sites of
recent/on-going star formation. Additional 22 GHz observations of the Galactic
AGB star R Dor are described in Appendix A. The IRAS names of the targets are
listed in Table 2, together with the IRAS flux densities at 12 and 25 $\mu$m,
object classification, and whether OH maser emission has been detected (yes),
not detected (no), or not tried ( ) by Wood et al.\ (1986, 1992) or van Loon
et al.\ (1998a). The results for the targets (tentatively) detected in at
least one of the H$_2$O or SiO maser transitions are summarised in Table 3,
whilst upper limits (3-$\sigma$ noise levels) for the non detections are
summarised in Table 4.

%
%
\begin{figure}[tb]
\centerline{\psfig{figure=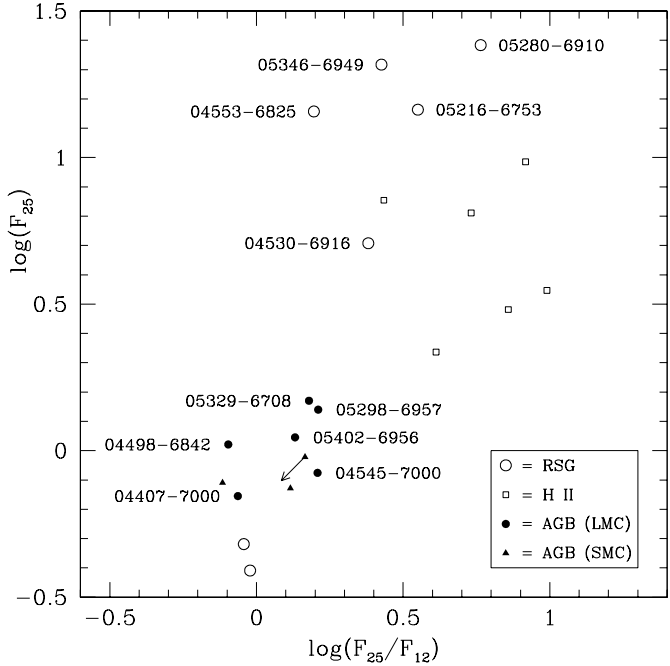,width=88mm}}
\caption[]{Diagram of 25 $\mu$m flux density versus 25 over 12 $\mu$m flux
density ratio for the targets of the maser search, distinguished according to
their object classes (RSGs, H {\sc ii} regions, AGB stars). The flux densities
of the SMC objects have been scaled by $+0.2$ dex to the distance of the LMC.}
\end{figure}

%
%
\begin{table*}
\caption[]{Peak flux densities ($\pm$ 1-$\sigma$, in Jy) for objects detected
(boldface) in a H$_2$O and/or SiO maser transition: H$_2$O 22 GHz with Parkes
(P; 1997 \& 2000) and Mopra (M), SiO 43 GHz with Parkes, SiO 86 GHz with Mopra
and SEST (S; HRS \& LRS), and SiO 129 GHz with SEST. Tentative detections are
between parentheses, 3-$\sigma$ upper limits are given for non detections.}
\begin{tabular}{l|lll|lllll}
\hline\hline
                            &
\multicolumn{3}{c|}{H$_2$O} &
\multicolumn{5}{c}{SiO}     \\
IRAS-PSC       &
P$_{22}$(1997) &
P$_{22}$(2000) &
M$_{22}$       &
P$_{43}$       &
M$_{86}$       &
S$_{86}$(HRS)  &
S$_{86}$(LRS)  &
S$_{129}$      \\
\hline
{\it LMC sources}              &
                               &
                               &
                               &
                               &
                               &
                               &
                               &
                               \\
04553$-$6825                   &
         {\bf 0.068$\pm$0.006} &
         {\bf 0.195$\pm$0.010} &
         {\bf 0.20$\pm$0.04}   &
       $<$0.67                 &
\llap{$<$}0.22                 &
         {\bf 0.150$\pm$0.025} &
         {\bf 0.072$\pm$0.011} &
\llap{$<$}0.15                 \\
05216$-$6753                   &
                               &
 \llap{(}0.046$\pm$0.020)      &
                               &
       $<$0.90                 &
                               &
\llap{$<$}0.28                 &
\llap{$<$}0.16                 &
                               \\
05280$-$6910                   &
                               &
         {\bf 0.137$\pm$0.010} &
 \llap{(}0.30$\pm$0.06)        &
       $<$1.5                  &
\llap{$<$}0.34                 &
 \llap{(}0.30$\pm$0.08)        &
\llap{$<$}0.22                 &
\llap{$<$}0.18                 \\
05329$-$6708                   &
 \llap{(}0.026$\pm$0.007)      &
\llap{$<$}0.027                &
\llap{$<$}0.09                 &
       $<$1.6                  &
                               &
 \llap{(}0.12$\pm$0.07)        &
\llap{$<$}0.14                 &
\llap{$<$}0.17                 \\
\hline
\end{tabular}
\end{table*}

%
%
\begin{table}
\caption[]{Upper limits (3-$\sigma$ noise levels, in Jy) for the non
detections: H$_2$O 22 GHz with Parkes (P, 2000) and Mopra (M), SiO 43 GHz with
Parkes, and SiO 86 GHz with Mopra.}
\begin{tabular}{l|ll|ll}
\hline\hline
                            &
\multicolumn{2}{c|}{H$_2$O} &
\multicolumn{2}{c}{SiO}     \\
IRAS-PSC       &
P$_{22}$(2000) &
M$_{22}$       &
P$_{43}$       &
M$_{86}$       \\
\hline
{\it SMC sources} &
                  &
                  &
                  &
                  \\
00483$-$7347 &
0.030        &
0.15         &
             &
             \\
00486$-$7308 &
0.033        &
             &
             &
             \\
01074$-$7140 &
0.039        &
             &
             &
             \\
\hline
{\it LMC sources} &
                  &
                  &
                  &
                  \\
04407$-$7000 &
             &
0.09         &
1.5          &
0.35         \\
04491$-$6915 &
             &
             &
1.5          &
             \\
04498$-$6842 &
             &
             &
2.0          &
             \\
04514$-$6931 &
             &
             &
2.0          &
             \\
04530$-$6916 &
0.039        &
0.12         &
             &
             \\
04545$-$7000 &
0.027        &
             &
2.2          &
             \\
04546$-$6915 &
             &
             &
2.2          &
             \\
04553$-$6933 &
             &
             &
2.0          &
             \\
04571$-$6627 &
             &
             &
2.0          &
             \\
04581$-$7013 &
             &
             &
2.0          &
             \\
05198$-$6941 &
             &
             &
2.0          &
             \\
05298$-$6957$^\dagger$ &
0.024        &
0.17         &
1.2          &
0.46         \\
05325$-$6743 &
             &
             &
1.6          &
             \\
05346$-$6949 &
0.033        &
0.11         &
1.5          &
             \\
05402$-$6956 &
0.036        &
0.15         &
1.2          &
             \\
\hline
\end{tabular}\\
$^\dagger$\footnotesize{observed with SEST at 86 GHz (HRS \& LRS) and 129 GHz,
with upper limits of 0.29, 0.09 and 0.17 Jy, respectively.}
\end{table}

The IRAS mid-IR flux densities and flux density ratios often carry information
about the nature of the IR source (Fig.\ 1). The AGB stars amongst the targets
represent the most luminous AGB stars with the highest mass-loss rates
encountered in the MCs, with typical mid-IR flux densities of $\sim1$ Jy.
Because of their higher luminosity and sometimes also higher mass-loss rate,
RSGs may become somewhat redder and more than an order of magnitude brighter
at 25 $\mu$m than these extreme AGB stars. RSGs, due to their short lifetimes,
may still be associated with H {\sc ii} regions that usually have very red
mid-IR colours. Very young supergiants may be dust-enshrouded inside of an
ultracompact H {\sc ii} region with similarly red colours (see, for instance,
Persi et al.\ 1994). The individual targets are described below --- with a
general reference to Loup et al.\ 1997 --- together with the results from our
maser search.

\subsection{RSGs in the LMC}

\subsubsection{IRAS04530$-$6916}

%
%
\begin{figure}[tb]
\centerline{\psfig{figure=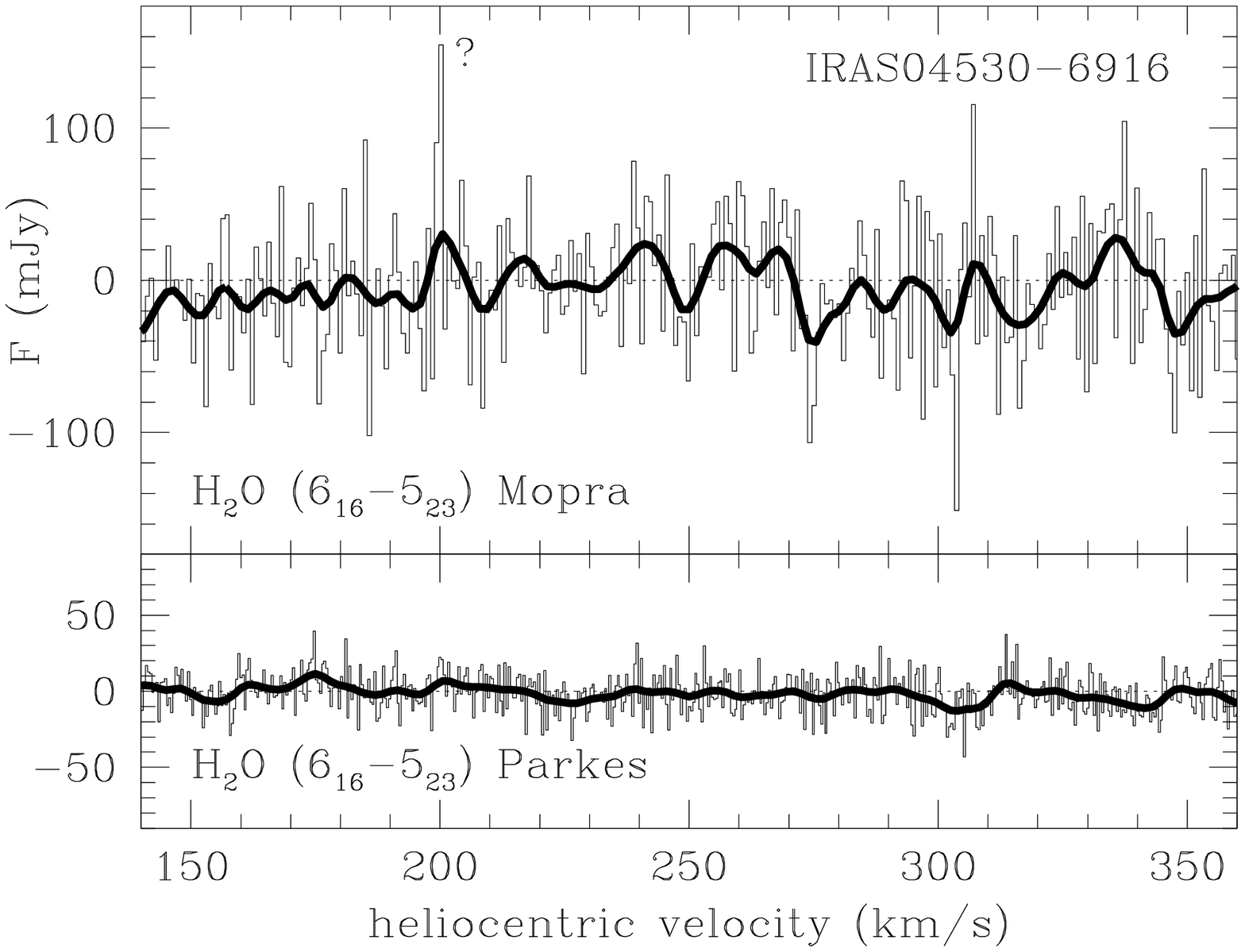,width=88mm}}
\caption[]{IRAS04530$-$6916: Mopra and Parkes (2000) 22 GHz spectra. The
velocities are heliocentric. The boldfaced curves are the spectra smoothed by
a gau{\ss}ian of $\sigma=2.0$ km s$^{-1}$. A possible H$_2$O maser peak at
$v_{\rm hel}\sim200$ km s$^{-1}$ in the Mopra spectrum could not be confirmed
at Parkes despite the much lower noise levels.}
\end{figure}

This source may be interpreted as a luminous RSG or AGB star, with a high
mass-loss rate of $\dot{M}\sim8\times10^{-4}$ M$_\odot$ yr$^{-1}$ (van Loon et
al.\ 1999b) and detected by IRAS even at 100 $\mu$m (Trams et al.\ 1999). It
is located in the H {\sc ii} region DEM L15. Variability in the near-IR is
reported by Wood et al.\ (1992), with a period of $\sim1260$ days, but the
amplitudes are only $\la0.4$ mag even in the J-band. An I-band spectrum of the
near-IR object associated with the IRAS source reveals an early-type
emission-line spectrum (Cioni et al., in preparation) adding some confusion
with regard to the nature of this object.

A narrow 3-$\sigma$ peak ($\sim150$ mJy) at $v_{\rm hel}\sim200$ km s$^{-1}$
in the Mopra 22 GHz spectrum could not be confirmed at Parkes at much lower
noise levels (Fig.\ 2), possibly due to temporal variability of the H$_2$O
maser emission (see, for instance, Persi et al.\ 1994) or to beamsize
differences if the source is at $\sim2^\prime$ from IRAS04530$-$6916. The
maser velocity would be rather different from the 21 cm H {\sc i} that peaks
at $v_{\rm hel}\sim260$ km s$^{-1}$ (Kim et al.\ 1999), but this is also the
case for some of the H$_2$O masers in the giant star forming region 30 Doradus
(van Loon \& Zijlstra 2000).

\subsubsection{IRAS04553$-$6933}

This supergiant is of spectral type M2 (WOH S71: Westerlund et al.\ 1981).

\subsubsection{IRAS04553$-$6825}

IRAS04553$-$6825 is a very luminous obscured RSG that has been extensively
discussed in the past (van Loon et al.\ 1999a,b and references therein). OH
maser emission at 1612 and 1665 MHz was discovered by Wood et al.\ (1986,
1992), SiO maser emission at 86 GHz by van Loon et al.\ (1996), and H$_2$O
maser emission at 22 GHz by van Loon et al.\ (1998b).

%
%
\begin{figure}[tb]
\centerline{\psfig{figure=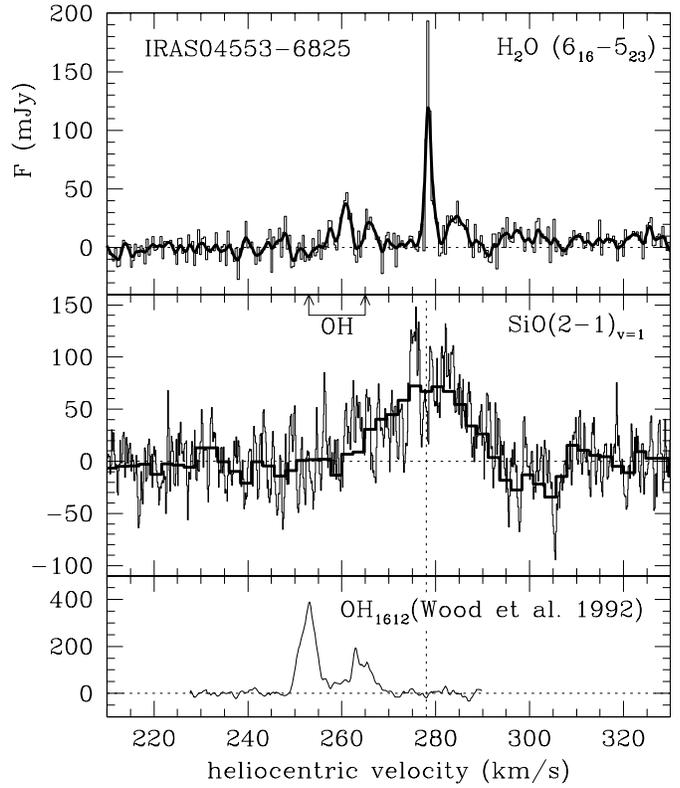,width=88mm}}
\caption[]{IRAS04553$-$6825: Spectra of the 22 GHz H$_2$O (top; Parkes 2000)
and 86 GHz SiO (middle; SEST) maser emission. The boldfaced curve in the upper
panel is the spectrum smoothed by a gau{\ss}ian of $\sigma=0.5$ km s$^{-1}$,
whilst in the middle panel it is the LRS spectrum. The velocities are
heliocentric. Also indicated are the velocities of the 1612 MHz OH maser peaks
(arrows) and the velocity of the H$_2$O maser peak (vertical dotted line). The
1612 MHz spectrum of Wood et al.\ (1992) is plotted in the lower panel for
comparison.}
\end{figure}

The H$_2$O maser emission from IRAS04553$-$6825 was confirmed with Mopra, and
re-observed with Parkes on two occasions in the 2000 run. The (integration
time \& system temperature weighted) average of the latter is presented in
Fig.\ 3 (top panel). The integrated flux of the H$_2$O maser emission between
255 and 307 km s$^{-1}$ equals 0.65 Jy km s$^{-1}$, corresponding to a photon
flux of $1.0\times10^{45}$ s$^{-1}$. The very narrow main peak is brighter in
the 2000 run at least partly due to the higher spectral resolution, and may,
in fact, still be unresolved. The double peak at blue-shifted velocities is
intriguing, as the shape closely resembles that of the 1612 MHz OH emission
profile (Wood et al.\ 1992) but at slightly smaller velocity displacements of
$-13$ and $-18$ km s$^{-1}$ with respect to the main H$_2$O maser peak at
$v_{\rm hel}=278.5$ km s$^{-1}$, instead of $-14$ and $-26$ km s$^{-1}$. This
strongly suggests that the blue-shifted H$_2$O maser peaks and the OH maser
peaks are formed in related but displaced regions in the CSE outflow. The
blue-shifted H$_2$O maser emission was not noticed in the discovery spectrum
(van Loon et al.\ 1998b) but can be recovered in that data {\it a posteriori}.
The additional red-shifted H$_2$O maser peak perhaps arises from some local
density enhancement or increased coherent path length in part of the CSE at
the far side of the star.

The composite spectrum of all SEST observations at 86 GHz of IRAS04553$-$6825
is presented in Fig.\ 3 (middle panel). The HRS spectrum is overlaid by the
LRS spectrum. The quality of the individual spectra is insufficient for a
detailed study of spectral changes with pulsational phase: no variability more
than a factor two is seen in the peak or integrated flux density. The SiO
maser properties have been listed in van Loon et al.\ (1998b). SiO maser
emission can be traced between velocities of $\sim260$ and 290 km s$^{-1}$,
and there is evidence for the peak of the emission to be split into two peaks
$\sim8$ km s$^{-1}$ apart. The shoulder of SiO maser emission between
$\sim260$ and 266 km s$^{-1}$ may be related to the blue-shifted H$_2$O maser
peaks. The velocity of the main H$_2$O maser peak falls right in the centre of
the double-peaked SiO emission.

Two lines at 298 and 305 km s$^{-1}$ are detected apparently in absorption
(Fig.\ 3, middle panel), and they are also visible at several individual
epochs. SiO masing lines have never been seen in absorption in Galactic
sources for good reasons: the excitation temperature is $\sim1800$ K. The
features may correspond to SiO emission in the off-source beam. It is hard to
trace back this source as the beamswitching was done in azimuth and hence the
off-source beam swept across the sky during each observing session.

\subsubsection{IRAS04581$-$7013}

This is a variable star (HV2255) of spectral type M4.

\subsubsection{IRAS05216$-$6753}

This object is very bright in the IR, with possibly an early-type supergiant
underlying its CSE (Zijlstra et al.\ 1996). Marginal variability is detected
by Wood et al.\ (1992), who suggest it may be a proto-PN or post-RSG object
and who note the similarity with IRAS04530$-$6916.

%
%
\begin{figure}[b]
\centerline{\psfig{figure=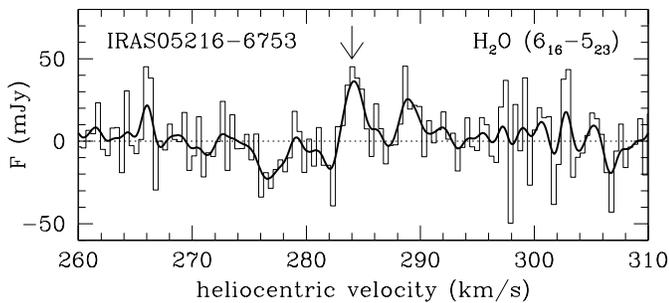,width=88mm}}
\caption[]{IRAS05216$-$6753: Spectrum of the tentative detection of 22 GHz
H$_2$O maser emission (indicated by the arrow). The velocities are
heliocentric. The boldfaced curve is the spectrum smoothed by a gau{\ss}ian of
$\sigma=0.5$ km s$^{-1}$.}
\end{figure}

The 22 GHz spectrum with Parkes shows a peak of $\sim45$ mJy --- a 3 to
4-$\sigma$ level after some smoothing --- at $v_{\rm hel}\sim284$ km s$^{-1}$
(Fig.\ 4). The peak has a FWHM of $\sim3$ km s$^{-1}$. The integrated flux of
the emission is $\sim0.5$ Jy km s$^{-1}$, corresponding to a photon flux of
$\sim8\times10^{44}$ s$^{-1}$. There may be a secondary peak at 289 km
s$^{-1}$.

\subsection{IRAS05280$-$6910}

The 1665 MHz OH maser attributed to IRAS05280$-$6910 is located at the centre
of the cluster NGC1984 (Wood et al.\ 1992). The IRAS 12, 25 and 60 $\mu$m flux
densities suggest a very cool and luminous CSE, possibly a post-RSG object.
The proposed near-IR counterpart is with $K=8.19$ mag very bright, but with
$(J-K)=1.01$ not very red (Wood et al.\ 1992). Within the Parkes beam of
$\sim13^\prime$ at OH frequencies are three luminous late type variable stars
of $V\sim13$ and $I\sim10$ mag, a PN (SMP LMC 64), several IR objects, the H
{\sc ii} region DEM L198 at $4.2^\prime$ W, and the cluster NGC1994
(IRAS05287$-$6910) at $\sim4^\prime$ E. It may well be that only the 1612 MHz
emission arises from the CSE of an evolved star in the region of NGC1984,
whilst the 1665 MHz emission arises from the ISM.

%
%
\begin{figure}[tb]
\centerline{\psfig{figure=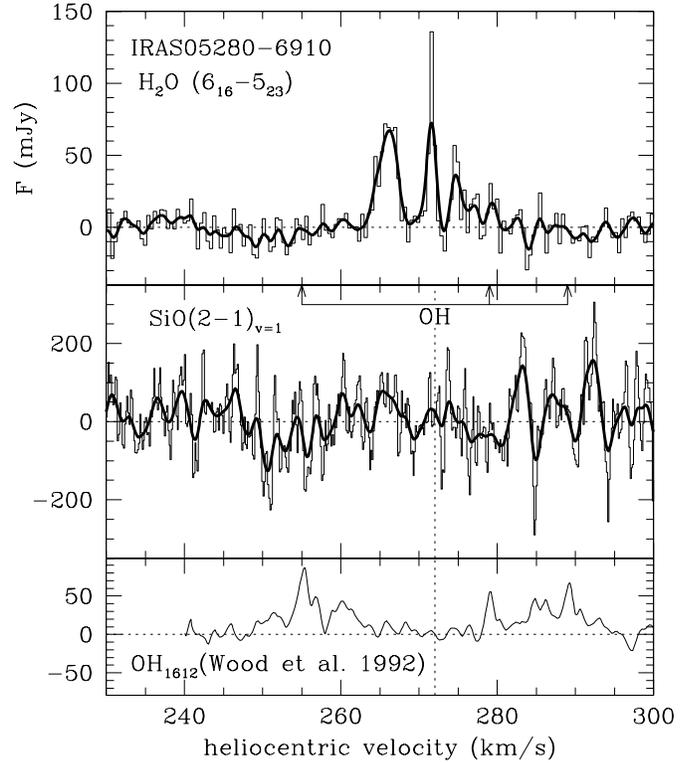,width=88mm}}
\caption[]{IRAS05280$-$6910: Spectra of the 22 GHz H$_2$O (top; Parkes 2000)
and 86 GHz SiO (middle; SEST) maser emission. The boldfaced curves are the
spectra smoothed by a gau{\ss}ian of $\sigma=0.5$ km s$^{-1}$. The velocities
are heliocentric. Also indicated are the velocities of the 1612 MHz OH maser
peaks (arrows) and the velocity of the H$_2$O maser peak (vertical dotted
line). The 1612 MHz spectrum of Wood et al.\ (1992) is plotted in the lower
panel for comparison.}
\end{figure}

Bright and complex H$_2$O maser emission was detected with Parkes (Fig.\ 5,
top panel). The very narrow and probably unresolved main peak is at $v_{\rm
hel}=272$ km s$^{-1}$, which also defines the centroid velocity of the OH 1612
MHz maser emission as observed by Wood et al.\ (1992). This confirms the
cluster object IRAS05280$-$6910 to be the source of the 1612 MHz OH maser
emission. A rather broad and bright additional peak is blue-shifted by $\sim5$
km s$^{-1}$, and fainter and possibly sub-structured emission is seen at
red-shifted velocities of $\sim3$ to 8 km s$^{-1}$ with respect to the main
peak. The integrated flux of the H$_2$O maser emission between 260 and 284 km
s$^{-1}$ equals 0.42 Jy km s$^{-1}$, corresponding to a photon flux of
$6.5\times10^{44}$ s$^{-1}$, not much less than IRAS04553$-$6825. Both the
main and the blue-shifted additional peaks can be recovered {\it a posteriori}
in the Mopra 22 GHz data of IRAS05280$-$6910.

The 86 GHz SiO maser spectrum of IRAS05280$-$6910 is presented in Fig.\ 5
(middle panel). A constant baseline offset of 2 mJy has been subtracted. The
strongest peak, at a heliocentric velocity of 292 km s$^{-1}$, has a peak flux
density of 0.30 Jy ($=3.8\sigma$). There is another ($2.8\sigma$) peak at a
velocity of 284 km s$^{-1}$, and a hint of faint emission at $v_{\rm
hel}\sim265$ km s$^{-1}$ (at which also H$_2$O maser emission is seen). The
integrated flux of the emission around 265, 284 and 292 km s$^{-1}$ equals
$\sim0.8$ Jy km s$^{-1}$, corresponding to a photon flux of $1.2\times10^{45}$
s$^{-1}$. The SiO detection is rather uncertain, though, and might still be
(strong) noise.

\subsubsection{IRAS05346$-$6949}

This is thought to be a very luminous and highly obscured RSG (Elias et al.\
1986), but this is still uncertain. Its IRAS-LRS spectrum is odd in that it
shows a flat continuum with peculiar features (Kwok et al.\ 1997).

\subsection{AGB stars in the LMC}

The AGB stars amongst the LMC targets are oxygen rich, very luminous, variable
on long timescales ($\sim10^3$ days) and severely dust-enshrouded (Wood et
al.\ 1992; van Loon et al.\ 1998a, 1999a,b). For some, mid-IR spectra have
been obtained by Trams et al.\ (1999), showing the 10 $\mu$m silicate dust
feature in absorption with some emission wings remaining (see also Groenewegen
et al.\ 1995 and Zijlstra et al.\ 1996 for IRAS05329$-$6708).

These stars comprise all known AGB sources of OH maser emission in the LMC,
except the bluest amongst them, IRAS04498$-$6842, that has not (yet) been
detected at 1612 MHz. The only typical saturated double-peaked OH maser
emission profile is found in IRAS05298$-$6957. IRAS04407$-$7000 is single
peaked at 1612 MHz (van Loon et al.\ 1998a), and consequently no expansion
velocity of the CSE can be determined, whilst the other emission profiles are
rather complex and/or faint.

\subsubsection{IRAS04545$-$7000}

This object is located in a small double association of stars (\#200 \& 201 in
Bica et al.\ 1999), with a cluster at $\sim2^\prime$.

\subsubsection{IRAS05298$-$6957}

This star is in a cluster containing a carbon star and has an initial mass of
$M_{\rm initial}\sim4$ M$_\odot$ (van Loon et al.\ 2000).

\subsubsection{IRAS05329$-$6708}

The brightest AGB star in the LMC at 25 $\mu$m, it was first identified with
the optically visible semi-regular variable TRM060 (Reid et al.\ 1990) before
the IRAS source was identified in the near-IR at $\sim26^{\prime\prime}$ S of
TRM060 by Wood et al.\ (1992). The region around IRAS05329$-$6708 is very
crowded, with a PN (TRM058 = LI-LMC1280) at $2.2^\prime$ SW, and an M4 Iab RSG
(HV5933 = TRM063 = IRAS05334$-$6706) at $4.2^\prime$ NE. The latter has IRAS
12 and 25 $\mu$m flux densities of 0.30 and 0.21 Jy, respectively.

%
%
\begin{figure}[b]
\centerline{\psfig{figure=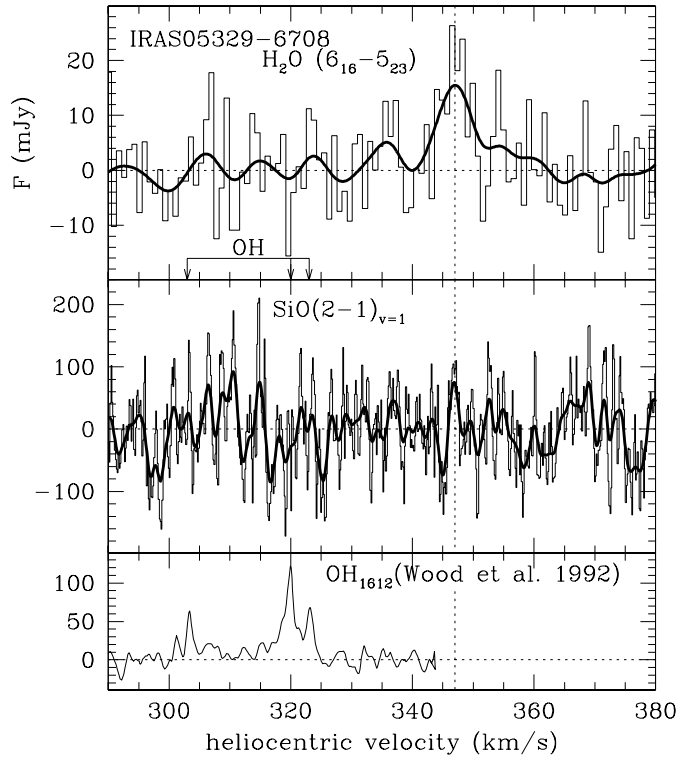,width=88mm}}
\caption[]{IRAS05329$-$6708: Spectra of the tentative detections of 22 GHz
H$_2$O (top; Parkes 1997) and 86 GHz SiO (middle; SEST) maser emission. The
boldfaced curves are the spectra smoothed by a gau{\ss}ian of $\sigma=2$ and
0.5 km s$^{-1}$, respectively. The velocities are heliocentric. Also indicated
are the velocities of the strongest 1612 MHz OH maser peaks (arrows), and
heliocentric velocity 347 km s$^{-1}$ (vertical dotted line). The 1612 MHz
spectrum of Wood et al.\ (1992) is plotted in the lower panel for comparison.}
\end{figure}

H$_2$O maser emission was tentatively detected (Fig.\ 6, top panel) at 22 GHz
at Parkes (1997 run), centred at $v_{\rm hel}=347$ km s$^{-1}$. The peak has a
FWHM of $\sim5$ km s$^{-1}$. The integrated flux of the emission is 0.11 Jy km
s$^{-1}$, corresponding to a photon flux of $1.7\times10^{44}$ s$^{-1}$. It is
not certain whether the 22 GHz peak corresponds to the stellar velocity, or
whether the OH and H$_2$O masers refer to the same source. The 86 GHz SiO
maser spectrum (Fig.\ 6, middle panel) is noisy, but there is positive signal
(integrated flux $\sim0.11$ Jy km s$^{-1}$) within 1 km s$^{-1}$ of the H$_2$O
maser peak, and some spikes within the velocity range of the OH emission.

\subsubsection{IRAS05402$-$6956}

This star is $\sim3^\prime$ W of an H {\sc ii} region (DEM L275 \& 277).

\subsection{AGB stars in the SMC}

No H$_2$O maser emission could be positively detected. To date, no
circumstellar maser emission has been detected from any source in the SMC.

\subsubsection{IRAS00483$-$7347}

This star is a late-M type LPV of $M_{\rm bol}\sim-7$ mag (Wood et al.\ 1992;
Groenewegen \& Blommaert 1998). Castilho et al.\ (1998) measure some
Li-enhancement. Their data suggest that the stellar photosphere is only mildly
metal-poor --- quite surprising for a star in the SMC. An H {\sc i} shell
(\#157 in Staveley-Smith et al.\ 1997) is located at $\sim1.5^\prime$ and
$v_{\rm hel}\sim156$ km s$^{-1}$.

%
%
\begin{figure}[b]
\centerline{\psfig{figure=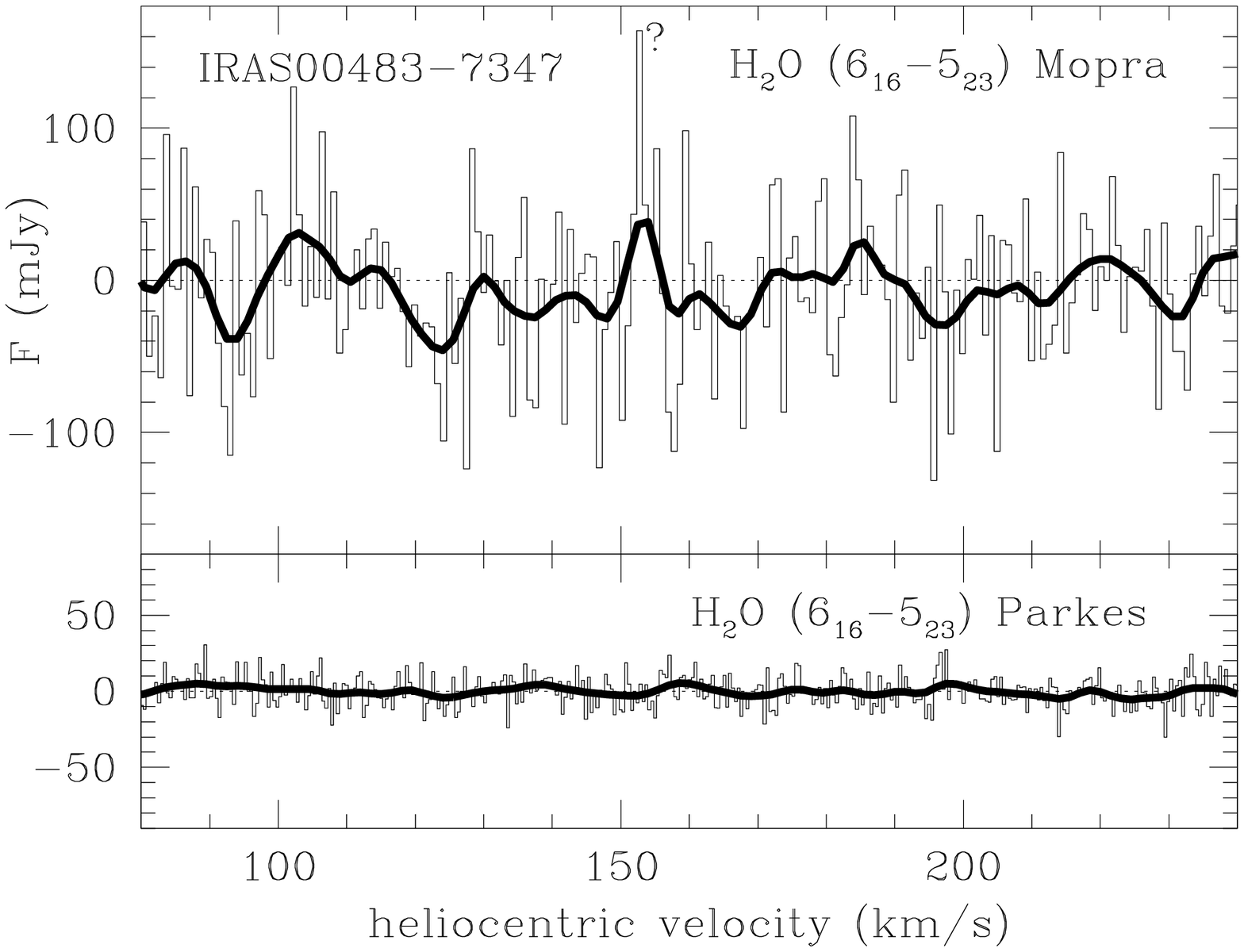,width=88mm}}
\caption[]{IRAS00483$-$7347: Mopra and Parkes (2000) 22 GHz spectra. The
velocities are heliocentric. The boldfaced curves are the spectra smoothed by
a gau{\ss}ian of $\sigma=2.0$ km s$^{-1}$. A possible H$_2$O maser peak at
$v_{\rm hel}\sim153$ km s$^{-1}$ in the Mopra spectrum could not be confirmed
at Parkes despite the much lower noise levels.}
\end{figure}

A 3-$\sigma$ peak ($\sim150$ mJy) at $v_{\rm hel}\sim153$ km s$^{-1}$ in the
22 GHz spectrum with Mopra could not be confirmed with Parkes (Fig.\ 7). It
would coincide with the maximum optical depth of the self-absorbed 21 cm H
{\sc i} (ATCA+Parkes) at $v_{\rm hel}\sim157$ km s$^{-1}$ (Stanimirovi\'{c} et
al.\ 1999).

\subsubsection{IRAS00486$-$7308}

This source is in the IRAS-FSC rather than the IRAS-PSC. The mid-IR flux
densities were determined by Groenewegen \& Blommaert (1998). It coincides
with the extended source IRAS0048$-$731, which may be identified with the H
{\sc ii} region LHA 115-N 36 and an H {\sc i} shell (\#156 in Staveley-Smith
et al.\ 1997). The most plausible counterpart of the IRAS point source is
GM103, a luminous late-M type AGB star (Groenewegen \& Blommaert 1998) from
which 10 $\mu$m silicate dust emission was detected by Groenewegen et al.\
(1995). An emission-line star (\#368 in Meyssonnier \& Azzopardi 1993) and a
carbon star (\#323 in Rebeirot et al.\ 1993) are $\sim0.5^\prime$ away.

\subsubsection{IRAS01074$-$7140}

This is a luminous M5e-type Li-enhanced variable AGB star (Wood et al.\ 1983;
Whitelock et al.\ 1989; Smith et al.\ 1995; Zijlstra et al.\ 1996; van Loon et
al.\ 1998a; Groenewegen \& Blommaert 1998), located $\sim0.8^\prime$ from the
centre of an H {\sc i} shell (\#369 in Staveley-Smith et al.\ 1997).

\subsection{H {\sc ii} regions in the LMC}

No 43 GHz SiO masers were found in any of these objects, but at rather high
detection thresholds. No other radio observations were made.

\subsubsection{IRAS04491$-$6915}

This source is associated with the H {\sc ii} region DEM L2 (LHA 120-N77D).

\subsubsection{IRAS04514$-$6931}

The extremely red mid-IR colour suggests that this is a (compact) H {\sc ii}
region.

\subsubsection{IRAS04546$-$6915}

This source is associated with the Wolf Rayet star HD32014 in/and the cluster
NGC1748, situated in the complex H {\sc ii} region DEM L22.

\subsubsection{IRAS04571$-$6627}

This source is located in the open cluster IC2116 that contains other
(early-type) evolved stars and that is embedded in the H {\sc ii} region LHA
120-N 11A (Parker et al.\ 1992; Rosado et al.\ 1996).

\subsubsection{IRAS05198$-$6941}

This luminous IR object is in a region of high stellar and nebular density
(LHA 120-N 120: Laval et al.\ 1992). Loup et al.\ (1997) identify the IRAS
point source with the WC star HD35517 (\#559 in Bohannan \& Epps 1974; see
also Laval et al.\ 1994). Within $\sim1^\prime$ are a multiple system (IDS
05201$-$6945) containing a B0 Iab supergiant, and an emission-line star (\#560
in Bohannan \& Epps 1974).

\subsubsection{IRAS05325$-$6743}

This is associated with the H {\sc ii} region LHA 120-N 57A.

\section{Stellar and outflow velocities from circumstellar masers in the LMC}

%
%
\begin{table}[b]
\caption[]{Kinematics for the presently known circumstellar masers in the LMC.
Outflow and stellar velocities ($v_\infty$ and $v_\star$, in km s$^{-1}$) are
derived both from OH alone (half the separation of the main peaks in the blue-
and red-shifted emission components) and from a combination of OH and SiO
and/or H$_2$O (the separation between the main peak of the blue-shifted OH
emission and the central peak of the SiO and/or H$_2$O emission). Uncertain
values are within parentheses. Also given are the 21 cm H {\sc i} heliocentric
velocities from the maps in LR-1992 = Luks \& Rohlfs (1992).}
\begin{tabular}{l|rr|rr|c}
\hline\hline
IRAS-PSC                              &
\multicolumn{2}{|c}{OH alone}         &
\multicolumn{2}{|c|}{OH + SiO/H$_2$O} &
LR-1992                               \\
                                      &
$v_\infty$                            &
$v_\star$                             &
$v_\infty$                            &
$v_\star$                             &
$v_{\rm H I}$                         \\
\hline
04407$-$7000 &   -         & 239 &   -         &    -         & [250,255] \\
04545$-$7000 &  (8\rlap{)} & 266 &   -         &    -         & [245,250] \\
04553$-$6825 &   7         & 260 &  26         &  278         & [280,285] \\
05216$-$6753 &   -         &   - &   -         &  284         & [295,300] \\
05280$-$6910 &  17         & 272 &  17         &  272         & [270,275] \\
05298$-$6957 &  11         & 282 &   -         &    -         & [265,270] \\
05329$-$6708 &  11         & 312 & (44\rlap{)} & (347\rlap{)} & [300,305] \\
05402$-$6956 &  11         & 272 &   -         &    -         & [260,265] \\
\hline
\end{tabular}
\end{table}

The presently available data on the stellar and outflow velocities of OH/IR
stars in the MCs are summarised in Table 5. Literature values are largely
based on the OH maser emission profile alone. The listed expansion velocities
in that case are derived from the separation of the strongest peaks in the
blue- and red-shifted emission components, respectively. Some authors use
instead the maximum extension of the emission (rather than the peaks) as a
measure for the expansion velocity (e.g.\ Zijlstra et al.\ 1996). Groenewegen
et al.\ (1998) compare expansion velocities derived from the OH peaks with
those derived from the width of the thermal CO emission and find that the
former are on average smaller by a factor 1.12 (see also Lewis 1991). In some
cases for which additional SiO and/or H$_2$O maser emission was discovered,
the OH masers appear to represent the part of the CSE in front of the star.
Hence the outflow velocities may have been severely underestimated. Revised
values are listed, using the OH peak velocities in relation to other maser
peaks. The interpretation of the maser emission from IRAS05329$-$6708, and
hence its stellar and expansion velocities, is rather uncertain.

\subsection{Stellar velocities and local H {\sc i} velocities}

The velocities of the OH/IR stars may be compared with the velocity of the
local ISM. Luks \& Rohlfs (1992) present a low spatial resolution map of 21 cm
H {\sc i} velocities for the LMC disk, from which approximate heliocentric
velocities are listed in Table 5. Kim et al.\ (1999) present a high spatial
resolution map of 21 cm H {\sc i} velocities, from which nine individual
spectra centred on and around the position of each OH/IR star are extracted
and averaged (Fig.\ 8).

%
%
\begin{figure}[tb]
\centerline{\psfig{figure=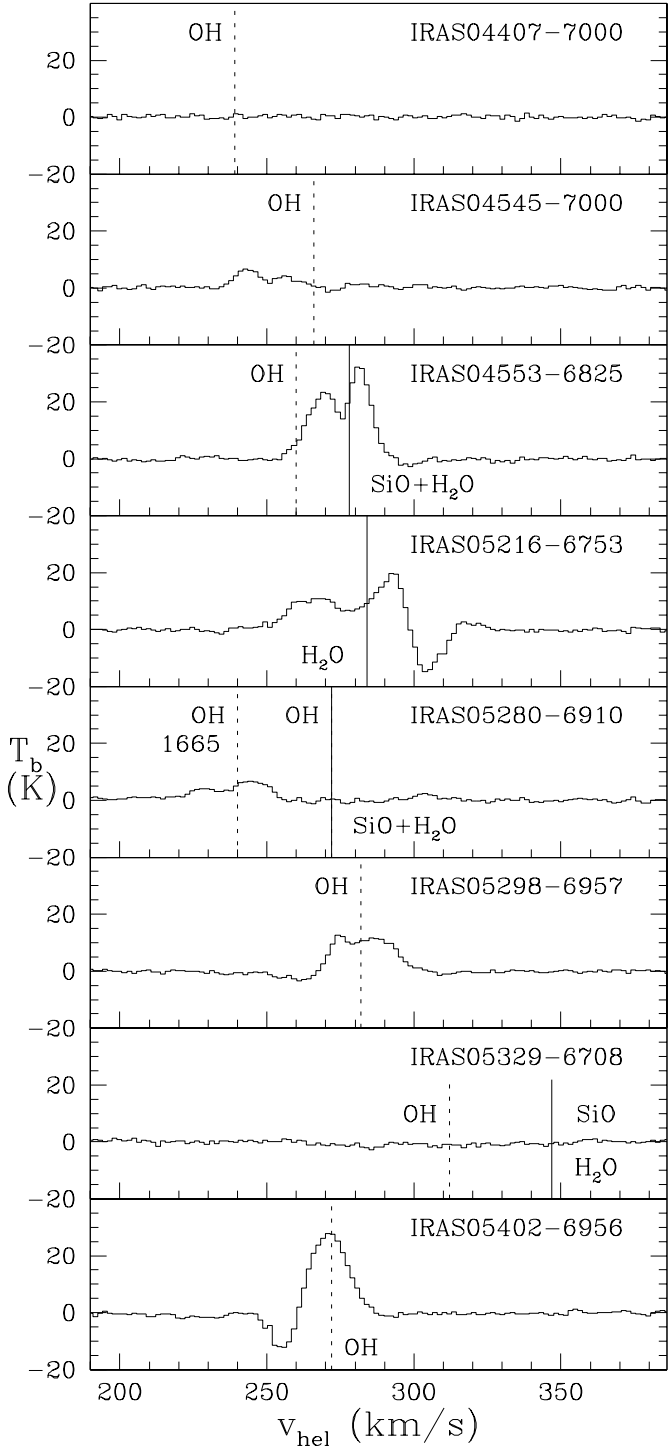,width=88mm}}
\caption[]{21 cm H {\sc i} spectra, constructed by averaging nine spectra from
Kim et al.\ (1999) centred on and around the position of each OH/IR star. The
stellar velocities are indicated as a vertical line (solid: SiO and/or
H$_2$O; dotted: 1612 MHz OH).}
\end{figure}

There is good agreement between the radio H {\sc i} emission and the H$\alpha$
emission around IRAS04553$-$6825 (Fig.\ B2). Both agree with its SiO and
H$_2$O maser velocities, as is to be expected for such a massive RSG. The
H$_2$O maser emission from IRAS05216$-$6753 provides for the first time a
radial stellar velocity for this heavily obscured object, consistent with the
local H {\sc i} velocity. The H$_2$O maser emission from IRAS05280$-$6910
peaks at the OH 1612 MHz centroid. Its stellar radial velocity deviates
significantly from the bulk of the H {\sc i} emission, with which the OH 1665
MHz emission is associated. The H$_2$O maser in IRAS05329$-$6708 suggests an
exceptionally high stellar velocity. It is seen projected on LMC4 (Meaburn
1980), a supergiant shell aged $\sim10^7$ yr, and peculiar velocities are not
surprising: high-velocity H$_2$O masers are also found in 30 Doradus (van Loon
\& Zijlstra 2000). Although most supergiants in the same (projected) region
have stellar velocities similar to the centre velocity of the OH profile
(Pr\'{e}vot et al.\ 1989), a few stars are found with radial velocities
$\sim350$ km s$^{-1}$ (Fehrenbach \& Duflot 1982).

Most OH/IR stars in the LMC follow the kinematics of the gas --- in particular
the AGB stars IRAS04545$-$7000, IRAS05298$-$6957 and IRAS05402$-$6956. As the
referee pointed out, this is remarkable because Galactic AGB stars show a
large velocity spread. The observation that the velocities of the OH/IR stars
in the LMC do not generally deviate (much) from the motion of the gas may be
due to the fact that the OH/IR stars are relatively massive. Even the AGB
stars amongst them probably have initial masses $M_{\rm initial}\ga4$
M$_\odot$, otherwise they would have become carbon stars (Wood 1998; van Loon
et al.\ 2000), and their ages are $t\la200$ Myr. If they have formed in the
disk of the LMC, they may still trace its rotation.

\subsection{Outflow kinematics in IRAS04553$-$6825}

The H$_2$O maser emission from IRAS04553$-$6825 (Fig.\ 2), with its double
blue-shifted emission peaks, reveals important information about the
acceleration of the outflow. In first instance only the central main peak of
H$_2$O maser emission was discovered (van Loon et al.\ 1998b). Although most
(circum-)stellar and maser properties of IRAS04553$-$6825 are virtually
identical to those of NML Cyg, a Galactic RSG (Morris \& Jura 1983; van Loon
et al.\ 1998b), the H$_2$O maser emission from NML Cyg (Richards et al.\ 1996)
nearly entirely arises from a bright double peaked structure at blue-shifted
velocities indicating outflow velocities in the H$_2$O masing region of
$\sim15$ km s$^{-1}$ --- a main peak of H$_2$O maser emission centred at the
stellar velocity has only occasionally been reported for NML Cyg. Comparison
with its OH maser emission indicative of outflow velocities of $\sim27$ km
s$^{-1}$ led to the interpretation of the H$_2$O masing region in NML Cyg to
be located in the accelerating part of the outflow where the matter has not
yet reached the local escape velocity (Richards et al.\ 1996). The detection
of similar blue-shifted H$_2$O maser emission from IRAS04553$-$6825 now for
the first time also allows us to measure the acceleration of the outflow for
this metal-poor RSG in the LMC.

The main H$_2$O maser peak in IRAS04553$-$6825 was interpreted by van Loon et
al.\ (1998b) either (i) to be radially beamed and hence indicate very low
outflow velocities of $\sim1$ km s$^{-1}$ in a dust-free inner CSE, or (ii) to
be tangentially beamed and centred at the stellar velocity. The detection of
the blue-shifted H$_2$O maser emission now strongly favours the latter. The
blue-shifted maser peaks suggest the emission is radially beamed and thus
measures the radial outflow velocity of the material expelled from the star,
in the region where the dust formation is thought to take place. Material is
accelerated from 18 km s$^{-1}$ in the H$_2$O masing z\^{o}ne to 26 km
s$^{-1}$ in the OH masing z\^{o}ne. The duplicity of both the H$_2$O and OH
blue-shifted emission suggests a second kinematic component in the CSE,
accelerating from 13 (H$_2$O) to 14 (OH) km s$^{-1}$. This slower component
may be closer to the star as suggested by the relatively stronger mainline OH
1665 MHz maser emission from that component (Wood et al.\ 1992). It is
remarkable that the OH 1665 MHz emission profile is broader than that of the
OH 1612 emission, a phenomenon that is attributed to either (or a combination
of) Zeeman broadening, clumpiness, velocity fluctuations or axi-symmetric
winds (Sivagnanam \& David 1999).

The kinematic data for the CSE of IRAS04553$-$6825 are very similar to the
kinematics in the CSE of NML Cyg, suggesting that the outflow kinematics
including the acceleration mechanism does not depend on metallicity. Other
data, however, seem to support theoretical expectations of lower velocities in
CSEs of lower metallicity (van Loon 2000). One way to reconcile both views is
if the distance to NML Cyg were $\sim1.3$ kpc rather than 2 kpc: the outflow
velocity scales with metallicity $Z$ and luminosity $L$ as $v_\infty \propto
\sqrt{Z} \sqrt[4]{L}$ (van Loon 2000) with $Z\sim0.02$ for NML Cyg (solar) and
$Z\sim0.008$ for IRAS04553$-$6825 (van Loon et al.\ 1998b and Appendix B).

The velocity range over which SiO maser emission is seen from IRAS04553$-$6825
covers the entire velocity range of the H$_2$O maser emission. If the strong
double peak of SiO maser emission is tangentially beamed then the velocity
separation might imply a rotational velocity component of the inner CSE. If
the double SiO peak is radially beamed, however, then it suggests moderate
outflow velocities in the dust-free inner CSE of $\sim4$ km s$^{-1}$. In any
case, the simultaneous presence of SiO emission over a large velocity extent
and strong discrete peaks of SiO emission indicate that the velocity field of
the inner CSE is highly complex.

Results from echelle spectroscopic observations of IRAS04553$-$6825 in the 0.6
to 0.9 $\mu$m region that generally support the kinematic picture for the CSE
of this star are described in Appendix B.

\subsection{Outflow kinematics in IRAS05280$-$6910}

The H$_2$O maser emission from IRAS05280$-$6910 essentially shows the same
features as seen in IRAS04553$-$6825, and thus carries the same potential for
analysing the kinematics of its CSE: the narrow main peak is interpreted as
tangentially beamed radiation centred at the stellar velocity, and the
blue-shifted emission (in the case of IRAS05280$-$6910 probably its
red-shifted counterpart is seen as well) can be explained as radially beamed
radiation indicating the radial outflow velocity in the H$_2$O masing region
of the CSE close to the region of dust formation. The outflow in the CSE of
IRAS05280$-$6910 is being accelerated from $\sim6$ km s$^{-1}$ in the H$_2$O
masing region to $\sim17$ km s$^{-1}$ in the outer CSE from where the OH 1612
MHz maser emission arises.

Double-peaked red-shifted SiO emission may have been detected, with equal
velocity separation but slightly larger receding velocities than the
red-shifted OH emission. This suggests that material in the dust-free inner
CSE may exhibit a wide range of velocities, possibly exceeding the final wind
velocity (Cernicharo et al.\ 1997).

\subsection{Outflow kinematics of Magellanic and Galactic circumstellar
masers}

Expansion velocities from the separation of the two OH peaks versus maximum
peak flux density are plotted in Fig.\ 9 for the OH/IR stars in the Galactic
Centre from Lindqvist et al.\ (1992a) and Sjouwerman et al.\ (1998), after
scaling their flux densities from the distance of the Galactic Centre to the
distance of the LMC. Also plotted are the expansion velocities for the OH/IR
stars in the LMC, as derived from the combination of all detected maser peaks.
Expansion velocities tend to be below average for the brightest OH sources.
For the LMC sources the statistics are very poor, with only one
well-determined expansion velocity for an AGB star (IRAS05298$-$6957)
suggestive of smaller expansion velocities at lower metallicity (see also Wood
et al.\ 1992; Zijlstra et al.\ 1996; van Loon 2000). The new maser data
presented here for IRAS05329$-$6708, however, suggest that this source may
exhibit an exceptionally large expansion velocity even when compared to the
expansion velocities of Galactic post-AGB objects (Zijlstra et al.\ 2000).
Disregarding IRAS04553$-$6825 and IRAS05329$-$6708, the average expansion
velocity of the remaining four LMC objects is $v\sim12\pm3$ km s$^{-1}$, which
may be compared to $v\sim14\pm2$ km s$^{-1}$ of the six brightest OH masers in
the Galactic Centre. Hence the expansion velocity of bright OH sources in the
LMC is $\sim20$\% lower than that of similarly bright OH masers in the
Galactic Centre, but the data is also consistent with no difference in
expansion velocity.

%
%
\begin{figure}[tb]
\centerline{\psfig{figure=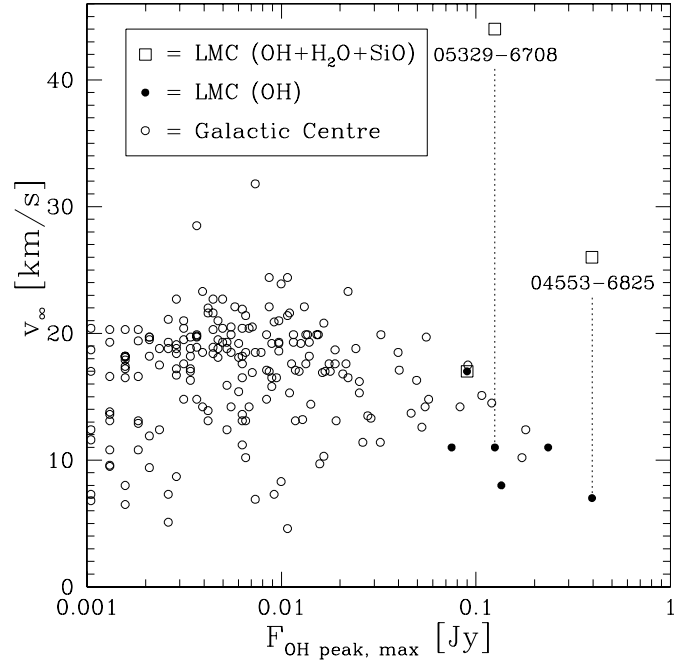,width=88mm}}
\caption[]{Expansion velocities of the circumstellar shells derived from the
separation of the OH maser peaks, versus the flux density of the brightest OH
maser peak (at the LMC distance). Expansion velocities derived from the
combination of all detected masers are given too for the LMC sources
(squares).}
\end{figure}

In the Galactic Centre, unlike the 13 SiO masers (Lindqvist et al.\ 1991) that
peak within a few km s$^{-1}$ of the mid-velocity of the OH peaks --- i.e.\
centred at the stellar velocity --- the H$_2$O detections (Lindqvist et al.\
1990) peak at or very near the OH peak velocities. This suggests that the wind
has already (nearly) reached its terminal velocity before leaving the H$_2$O
masing z\^{o}ne. In three out of four cases the blue-shifted H$_2$O peak is
(much) brighter than the red-shifted peak. H$_2$O masers in the LMC, however,
peak at the stellar velocity indicated by the centroid of the SiO maser
emission, rather than at (one of) the OH maser peak(s). Secondary H$_2$O peaks
suggest that the wind must still experience substantial acceleration after
leaving the H$_2$O masing z\^{o}ne. This may be understood by a less efficient
acceleration in the H$_2$O masing z\^{o}ne of low-metallicity CSEs, which
results in a stellar wind that is still being accelerated upon entering the OH
masing z\^{o}ne. This could cause multiple, thin OH masing shells to give rise
to the rather irregular OH emission profiles as observed in the LMC.

\subsection{Asymmetric emission profiles:\\
\hspace{7mm}non-spherical outflow or radiation transfer effects?}

All masers observed in the LMC are blue-asymmetric in the sense that the
masing material approaching Earth (after correcting for the stellar velocity
with respect to Earth) appears brighter than the receding matter (Fig.\ 10):
the OH, H$_2$O and SiO maser emission from IRAS04553$-$6825, the OH and H$_2$O
maser emission from IRAS05280$-$6910, the OH maser emission from all four AGB
stars detected by Wood et al.\ (1992) --- if interpreting IRAS05329$-$6708 in
combination with the tentative H$_2$O maser detection --- and possibly also
the single-peaked OH maser emission from IRAS04407$-$7000. Does this reflect
outflow complexity (Zijlstra et al.\ 2000), or is it merely due to the origin
and propagation of the amplified radiation field? Blue-asymmetric emission
profiles (Norris et al.\ 1984; Sivagnanam et al.\ 1990) arise if the
(radially-beamed) masers amplify stellar light or free-free emission from the
inner part of the CSE, rather than radiation from the dusty CSE, and/or the
receding maser spots are occulted by the star or an optically thick
free-electron reservoir.

%
%
\begin{figure}[tb]
\centerline{\psfig{figure=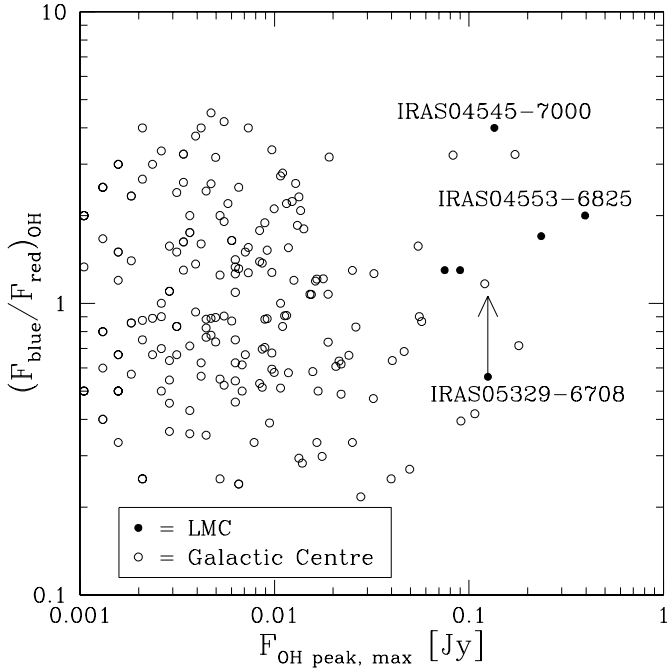,width=88mm}}
\caption[]{Ratio of the flux densities of the blue- and red-shifted OH maser
peaks, versus the flux density of the brightest OH maser peak (at the LMC
distance).}
\end{figure}

The blue asymmetry is expected to be more pronounced in RSGs such as
IRAS04553$-$6825 than in AGB stars such as IRAS05298$-$6957, because (i) at mm
wavelengths, the ratio of stellar light to CSE radiation is larger for RSGs
than for more compact AGB stars that have optically thicker CSEs (van Loon
2000), and (ii) RSGs are generally warmer than AGB stars and hence the
free-electron abundance is higher around RSGs. The blue asymmetry may be more
pronounced at lower metallicity, because (i) at mm wavelengths, the ratio of
stellar light to CSE radiation is smaller because of the lower dust-to-gas
ratio, and (ii) the free-electron abundance is higher due to the warmer
photospheres of low-metallicity stars.

\section{Photon fluxes of Magellanic and Galactic circumstellar masers}

\subsection{OH masers}

The OH masers in the LMC have large photon fluxes compared to what is typical
for the Galactic Centre (see Fig.\ 9), where the OH luminosity function peaks
at $2.5\times10^{43}$ s$^{-1}$ (Sjouwerman et al.\ 1998). The OH maser
emission from IRAS04553$-$6825 is $\sim100$ times brighter (and so is NML Cyg,
a very similar RSG in the Milky Way). The sixth brightest maser in the
Galactic Centre is equally bright as the sixth brightest maser in the LMC,
which suggests that a similar number of intermediate-age stars exist in the
Galactic Centre and in the LMC. Saturated OH masers have flux densities
$S_{\rm OH}\sim\frac{1}{4}{\times}S_{35 \mu{\rm m}}$. Although the detected OH
masers in the Magellanic AGB stars meet the saturation criterion within a
factor $\sim2$, the detected OH masers in the two Magellanic RSGs are an order
of magnitude below the saturation level.

The deepest search for OH sources in the LMC has been performed with a
detection limit of $\sim50$ mJy, and several single-peaked sources were found,
including IRAS04407$-$7000 (van Loon et al.\ 1998a). A flux density of 50 mJy
from the LMC corresponds to 2 Jy from the Galactic Centre. Of the 134 double
and 16 single-peaked OH sources in Lindqvist et al.\ (1992a), only 12 have
maximum flux densities $>2$ Jy. Out of these 12 sources, 8 would appear as a
single peak. Hence, when searching for OH masers close to the detection limit,
one is likely to find more single than double-peaked sources.

Many of the 1612 MHz peaks of van Loon et al.\ (1998a) could not be identified
with IR-bright point sources associated with CSEs around evolved stars.
Interestingly, none of the Galactic Centre OH sources with maximum flux
densities $>2$ Jy were recovered in the near-IR by Wood et al.\ (1998), who
did identify a number of the fainter Galactic Centre OH sources with near-IR
objects. Only two of these bright OH sources are identified with IRAS point
sources (Appendix C).

Amongst the LPVs with known pulsation periods and bolometric magnitudes from
Wood et al.\ (1998) the fraction of stars with detected OH emission increases
with redder $(K-L)_0$: 0.22, 0.77, 0.90, 0.95 and 1.00 for colour bins
$({\leftarrow},1]$, $[1,2]$, $[2,3]$, $[3,4]$ and $[4,{\rightarrow})$ mag,
respectively. No OH masers are detected from LPVs with $P\la500$ d. The OH/IR
stars in the LMC all have periods $900{\la}P\la1300$ d, but their colours are
with $(K-L)_0\sim2$ mag not extremely red (Trams et al.\ 1999) due to the low
dust content as a result of a low metallicity.

\subsection{$H_2$O masers}

Lindqvist et al.\ (1990) detected 22 GHz H$_2$O maser emission from 4 out of
33 OH sources in the Galactic Centre. Their detections typically have peak
flux densities of 0.5 to 1 Jy, corresponding to $\sim20$ mJy if the sources
were at the distance of the LMC. The 3-$\sigma$ upper limits for their
non-detections correspond to $\sim10$ mJy at the distance of the LMC, which is
comparable to the sensitivity of our search for H$_2$O maser emission from
Magellanic sources. Of the 13 Magellanic IRAS sources that were observed at 22
GHz, four were detected. Three of the IRAS-PSC identifications with OH masers
in the Galactic Centre (Table C1) were observed at 22 GHz: OH359.675$+$0.069
and OH359.946$-$0.047 were detected (both identified with near-IR LPVs), but
the bright OH maser OH359.762$+$0.120 (for which no near-IR counterpart has
been identified) was not detected. Lewis (1998) already remarked that 22 GHz
detection rates are always $<100$\% (namely, $\sim80$\%), and even some bright
OH sources are undetected at 22 GHz. He suggests that this may be due to the
absence of density enhancements (``blobs'') or an unfavourable orientation of
a bipolar geometry. Hence, not all Magellanic OH/IR stars should be expected
to exhibit (strong) H$_2$O masers. Significant mass loss is required for the
presence of circumstellar H$_2$O masers, though: AGB stars with modest mass
loss are far below the sensitivity of both the Magellanic and Galactic Centre
22 GHz surveys (see Appendix A).

The ratio of H$_2$O and OH photon fluxes is $\sim0.1$ to 1, and the ratio of
H$_2$O and OH peak flux densities is $\sim1$, both for the Galactic Centre as
well as the LMC masers. At lower metallicities the CSEs are optically thinner
and provide less self-shielding from the interstellar UV radiation field, and
hence H$_2$O may be expected to be dissociated over a larger extent of CSEs in
the LMC compared to those in the Galactic Centre (Huggins \& Glassgold 1982),
thus yielding relatively faint H$_2$O maser emission compared to the OH maser
emission. The fact that no difference is seen between these relative
intensities in the LMC and the Galactic Centre suggests that the effect of
stronger dissociation might be cancelled by the effect of longer coherent
paths throughout the more slowly accelerating wind in lower metallicity CSEs.

\subsection{SiO masers}

Lindqvist et al.\ (1991) detected 43 GHz SiO maser emission from 13 out of 31
OH sources in the Galactic Centre, peaking at $S_{43}\sim1$ Jy ($\sim30$ mJy
at the LMC). Upper limits for the photon rates of Galactic non-detections at
43 GHz were a few $10^{44}$ s$^{-1}$. The upper limits for LMC sources using
the Parkes dish are a few dozen times brighter and thus not very useful. None
of the Galactic Centre 43 GHz masers could be detected at 86 GHz down to
typical photon rates $\sim10^{44}$ s$^{-1}$. This is similar to the SEST
limits for LMC sources and thus explains the difficulty in detecting 86 GHz
masers in the LMC.

Two of the Galactic Centre OH masers with IRAS-PSC identifications (Table C1)
were observed at 43 GHz, and both were detected (OH359.675$+$0.069 and
OH359.762$+$0.120). Three Galactic Centre H$_2$O masers were observed at 43
GHz, of which two were detected (including OH359.675$+$0.069). In the LMC too,
the brightest (circumstellar) H$_2$O masers are also the brightest SiO masers.
However, some quite bright mid-IR objects could not be detected in any maser
transition --- e.g.\ IRAS05346$-$6949.

%
%
\begin{figure}[tb]
\centerline{\psfig{figure=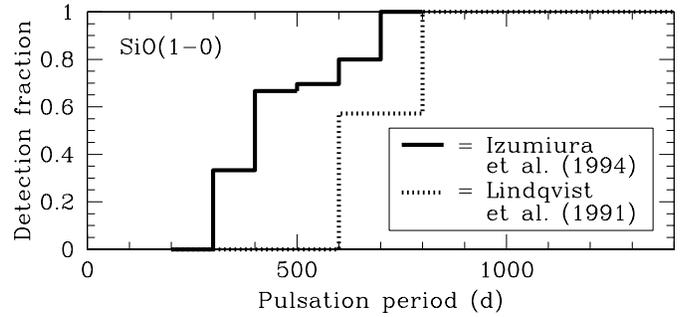,width=88mm}}
\caption[]{Detection fraction for SiO $J=1\rightarrow0$ masers versus
pulsation period, for the Galactic Bulge (solid: Izumiura et al.\ 1994) and
the Galactic Centre (dotted: Lindqvist et al.\ 1991).}
\end{figure}

SiO maser emission may be detected from IR objects with pulsation periods
$P\ga400$ d, and it is always present when $P\ga800$ d (Fig.\ 11). The OH/IR
stars in the LMC all have $P\ga900$ d and are thus expected to exhibit SiO
maser emission, probably not much below the sensitivity limits of our
searches. With $(K-L)_0\sim2$ mag their 86 GHz SiO maser emission is expected
to be similarly bright as their 43 GHz masers (Nyman et al.\ 1993). Lindqvist
et al.\ (1991) find that for Miras --- probably the progenitors of OH/IR stars
--- the SiO masers can often be $10^2$ times as intense as the OH masers. OH
luminosities scale with 35 $\mu$m luminosities, and OH outflow velocities
reach a maximum for OH/IR stars (Sivagnanam et al.\ 1989), indicating that
OH/IR stars experience stronger mass outflows than Miras. The ratio of SiO and
OH photon fluxes (or peak flux densities) is about unity both for the Galactic
Centre and LMC sources, which suggests that also in the LMC the brightest OH
masers are already experiencing heavy mass loss for some time.

%
%
\begin{table}
\caption[]{IR magnitudes and their variability for LMC masers (Elias et al.\
1986; Wood et al.\ 1992; Zijlstra et al.\ 1996; van Loon et al.\ 1998a) and
Galactic masers (Harvey et al.\ 1974; Le Bertre 1993), and SiO$_{\rm
v=1}(J=2\rightarrow1)$ peak flux densities (in Jy) for LMC masers (this work)
and Galactic masers (Nyman \& Olofsson 1985; Alcolea et al.\ 1990; Haikala
1990; Le Bertre \& Nyman 1990; Nyman et al.\ 1993; Haikala et al.\ 1994;
Bujarrabal et al.\ 1996; Cernicharo et al.\ 1997; Gonz\'{a}lez-Alfonso et al.\
1998; Herpin et al.\ 1998; Nyman et al.\ 1998).}
\begin{tabular}{lrrrrr}
\hline\hline
Star             &
$K$              &
${\Delta}K$      &
$N$              &
${\Delta}N$      &
SiO              \\
\hline
{\it LMC masers} &
                 &
                 &
                 &
                 &
                 \\
IRAS04407$-$7000 &
9.05             &
1.40             &
5.03             &
                 &
$<0.35$          \\
IRAS04553$-$6825 &
6.99             &
0.30             &
1.74             &
                 &
0.15             \\
IRAS05216$-$6753 &
10.38            &
0.20             &
2.43             &
                 &
$<0.28$          \\
IRAS05280$-$6910 &
8.19             &
                 &
2.41             &
                 &
(0.30\rlap{)}    \\
IRAS05298$-$6957 &
10.29            &
2.00             &
4.14             &
                 &
$<0.29$          \\
IRAS05329$-$6708 &
9.87             &
1.90             &
3.81             &
                 &
(0.12\rlap{)}    \\
\hline
{\it Milky Way masers} &
                       &
                       &
                       &
                       &
                       \\
IRAS00193$-$4033 &
3.28             &
1.30             &
$-1.61$          &
1.04             &
23               \\
IRAS22231$-$4529 &
3.11             &
1.09             &
$-0.55$          &
0.82             &
12               \\
IRC$-$30100      &
1.79             &
0.27             &
                 &
                 &
34               \\
IRC$-$30050      &
2.55             &
0.67             &
                 &
                 &
11               \\
IRC$-$20540      &
2.21             &
0.99             &
                 &
                 &
10               \\
IRC$-$20197      &
2.49             &
1.26             &
$-2.42$          &
1.30             &
45               \\
IRC$-$10529      &
2.33             &
1.50             &
$-3.31$          &
2.85             &
53               \\
IRC$+$10011      &
2.47             &
1.79             &
$-3.00$          &
1.70             &
66               \\
IRC$+$10523      &
1.85             &
1.07             &
                 &
                 &
10               \\
NML Cyg          &
0.60             &
0.51             &
$-5.30$          &
0.51             &
42               \\
NML Tau          &
$-0.68$          &
1.09             &
                 &
                 &
350              \\
OH02.60$-$0.4    &
3.42             &
1.52             &
                 &
                 &
27               \\
OH26.5$+$0.6     &
8.55             &
3.27             &
                 &
                 &
5                \\
OH285.05$+$0.07  &
5.32             &
1.50             &
$-0.12$          &
1.27             &
4                \\
OH286.50$+$0.06  &
5.27             &
1.41             &
$-0.43$          &
1.27             &
12               \\
OH300.93$-$0.03  &
4.95             &
1.20             &
                 &
                 &
2                \\
OH315.22$+$0.01  &
5.87             &
1.35             &
                 &
                 &
4                \\
OH341.12$-$0.01  &
6.04             &
1.16             &
                 &
                 &
2                \\
OH342.01$+$0.25  &
3.90             &
1.21             &
                 &
                 &
11               \\
OH344.83$-$1.67  &
7.02             &
1.28             &
                 &
                 &
2                \\
OH346.86$-$0.18  &
4.01             &
1.53             &
                 &
                 &
3                \\
OH349.18$+$0.20  &
8.34             &
1.64             &
                 &
                 &
2                \\
OH358.16$+$0.50  &
3.08             &
1.88             &
                 &
                 &
10               \\
R Aql            &
$-0.70$          &
0.58             &
$-2.40$          &
0.51             &
37               \\
R Aqr            &
$-1.09$          &
0.81             &
$-3.58$          &
0.70             &
140              \\
RR Aql           &
0.50             &
1.12             &
$-2.40$          &
1.08             &
29               \\
S Col            &
1.63             &
0.57             &
0.02             &
0.29             &
3                \\
S CrB            &
0.10             &
0.81             &
$-2.50$          &
0.58             &
100              \\
U Her            &
$-0.20$          &
0.81             &
$-2.50$          &
0.51             &
140              \\
U Ori            &
$-0.60$          &
0.90             &
$-2.70$          &
0.58             &
110              \\
VX Sgr           &
0.20             &
1.30             &
$-4.20$          &
0.90             &
280              \\
VY CMa           &
$-0.70$          &
0.15             &
$-5.90$          &
0.15             &
1700             \\
W Hya            &
$-3.10$          &
0.51             &
$-5.00$          &
0.28             &
680              \\
\hline
\end{tabular}
\end{table}

Alcolea et al.\ (1990) found a clear correlation between the visual amplitude
of pulsation and the pumping efficiency of the SiO masers. For our obscured
LMC sources no visual photometric monitoring data is available. It is known,
however, that the SiO maser emission correlates with the IR lightcurve (Nyman
\& Olofsson 1986; Alcolea et al.\ 1999), and hence we investigate here whether
the SiO maser intensity correlates with the IR amplitude. In Table 6 IR
photometric variability data is summarised for relatively nearby stars in the
Milky Way that were monitored by Harvey et al.\ (1974) and Le Bertre (1993).
The latter source, if available, is preferred because it is more modern by
nearly two decades. Where only narrow-band $N_{1,2,3}$ was obtained the
broad-band $N$ is approximated by an unweighted average. SiO$_{\rm
v=1}(J=2\rightarrow1)$ peak flux densities were compiled from recent
literature and averaged if more data was available. For the LMC sources
similar data is compiled, but at 10 $\mu$m too few epochs of measurements
exist to allow an estimate of the amplitude.

%
%
\begin{figure}[tb]
\centerline{\psfig{figure=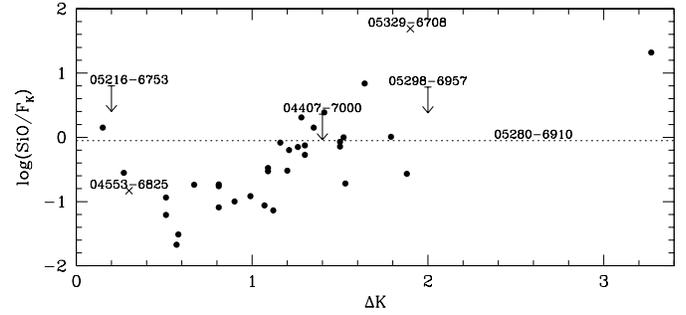,width=88mm}}
\caption[]{Flux density ratio of SiO$_{\rm v=1}(J=2\rightarrow1)$ peak and
K-band, versus K-band amplitude for Galactic (solid dots) and LMC masers from
Table 6.}
\end{figure}

The flux density ratio of the peak of the SiO$_{\rm v=1}(J=2\rightarrow1)$ and
the stellar flux at 2.2 $\mu$m increases for larger 2.2 $\mu$m amplitudes
(Fig.\ 12). This means that larger amplitudes cause relatively stronger maser
pumping, probably because of the relatively stronger shocks travelling through
the inner dust-free part of the CSE. A similar correlation is found for other
near-IR amplitudes and for integrated fluxes of the SiO maser emission, but
their lower accuracy leads to more scatter. The one secure and two tentative
detections of SiO$_{\rm v=1}(J=2\rightarrow1)$ maser emission in the LMC, as
well as the three useful upper limits are all consistent with SiO$_{\rm
v=1}(J=2\rightarrow1)$ masers that are equally strong in the LMC and the Milky
Way. Fig.\ 12 suggests ${\Delta}K\sim1$ to 2 mag for IRAS05280$-$6910, but the
DENIS K-band magnitude is with 8.195 (Cioni et al.\ 2000) very similar to the
photometry of Wood et al.\ (1992).

Interestingly, the flux density ratio of the peak of the SiO$_{\rm
v=1}(J=2\rightarrow1)$ and the stellar flux at 10 $\mu$m (N-band) is virtually
independent of any near-IR amplitude (Fig.\ 13; plotted against the amplitude
at 10 $\mu$m), and shows rather little scatter. This indicates a close
connection between the shocks that pump the SiO maser, and the dust formation.
It also suggests that both the SiO luminosity and the dust heating have a
common origin, namely the stellar flux. Again, the LMC data (no 10 $\mu$m
amplitudes available) is roughly compatible with the same trend. Thus there is
no strong evidence for a difference in maser strength at LMC metallicities
compared to $\sim$solar metallicity.

%
%
\begin{figure}[tb]
\centerline{\psfig{figure=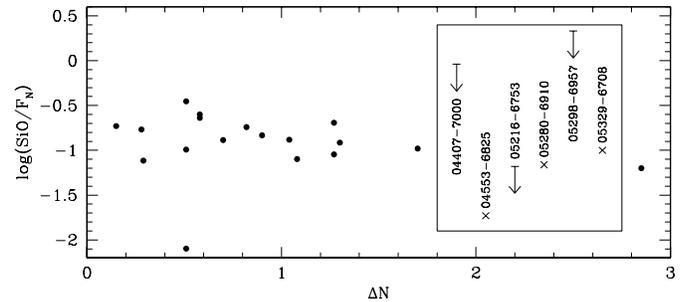,width=88mm}}
\caption[]{Same as Fig.\ 12, but for the N-band.}
\end{figure}

\section{Summary}

A five year effort to search for SiO and H$_2$O maser emission from
circumstellar envelopes in the LMC resulted in:\\
--- The secure detection of 86 GHz SiO$_{\rm v=1}(J=2\rightarrow1)$ and 22 GHz
H$_2$O $6_{16}\rightarrow5_{23}$ maser emission from the luminous red
supergiant IRAS04553$-$6825. The average SiO spectrum of 65 hours worth of
monitoring data shows unprecedented kinematic detail, probably indicating
outflow and/or turbulent motions with $v_{\rm SiO}\sim$ a few km s$^{-1}$ up
to $\sim18$ km s$^{-1}$ with respect to the stellar velocity. A new, improved
signal-to-noise H$_2$O spectrum also shows emission components at $v_{\rm
H2O}\la18$ km s$^{-1}$, that likely correspond to matter that is being further
accelerated before exhibiting OH masers at $v_{\rm OH}\sim26$ km s$^{-1}$.\\
--- The first detection of 22 GHz H$_2$O maser emission from the cluster
supergiant IRAS05280$-$6910. It consists of a central emission peak and
additional emission components indicating outflows with $v_{\rm H2O}\sim6$ km
s$^{-1}$, that are accelerated further before reaching the OH masing regions
that expand with $v_{\rm OH}\sim17$ km s$^{-1}$.\\
--- The tentative detection of 22 GHz H$_2$O maser emission from
IRAS05216$-$6753, probably a massive star inside an H {\sc ii} region. No
other masers have been detected.\\
--- The tentative detection of 22 GHz H$_2$O maser emission from the AGB star
IRAS05329$-$6708 is puzzling as it suggests an extremely fast outflow of
$v\sim44$ km s$^{-1}$.\\
--- Masers in the LMC are generally blue-asymmetric and/or single-peaked. We
propose that this may be due to the amplification of stellar and/or free-free
radiation, rather than dust emission. This may be more pronounced in low
metallicity envelopes due to the low dust content.\\
--- There is weak evidence for the expansion velocities in LMC objects to be
lower than of similar Galactic objects. The data is also consistent with no
difference in expansion velocities, due to the limited sample of LMC objects
with reliable estimates of the expansion velocity. The acceleration through
the CSE also seems to be slower in LMC objects, with the outflow velocity
increasing by a factor of two between the H$_2$O and OH masing z\^{o}nes.\\
--- SiO, H$_2$O and OH maser emission from circumstellar envelopes in the LMC
is found to be equally strong as from similar envelopes in the Milky Way. A
larger IR amplitude of variability leads to an increase in the flux density
ratio of the SiO$_{\rm v=1}(J=2\rightarrow1)$ peak and the 2.2 $\mu$m
continuum, but the flux density ratio of the SiO$_{\rm v=1}(J=2\rightarrow1)$
peak and the 10 $\mu$m spectral region is virtually a constant. This suggests
a close connection between the shocked SiO masing region and the dusty
outflow, which seems to be similar in the Milky Way and in the LMC.

Present-day facilities for observing SiO, H$_2$O and OH masers offer an
angular resolution and sensitivity capable of detecting only the very
brightest masers in the Magellanic Clouds. In the future, ALMA may provide
considerably larger samples of Magellanic SiO and H$_2$O masers, but for 1612
MHz OH masers no major improvement is envisaged. Detection of H$_2$O masers in
OH/IR stars in the Magellanic Clouds would, in principle, allow to derive the
total (gas+dust) mass-loss rates. Comparison of H$_2$O and OH deduced
expansion velocities yields the acceleration of the radiation-driven wind, if
the location of the H$_2$O and OH masers is known with sufficient accuracy. As
the distances and hence the luminosities for these stars are known, the
gas-to-dust ratios and total mass-loss rates are derived simultaneously
(Netzer \& Elitzur 1993).

\begin{acknowledgements}
We would like to thank the staff at the SEST, Parkes and Mopra observatories
for their kind and helpful support, and in particular Drs.\ Peter te Lintel
Hekkert, Marcus Price and Ian Stewart for help with the 22 GHz observations at
Parkes in August 1997. We also thank Dr.\ Roland Gredel for help with the NTT
observations at La Silla in January 1996. We acknowledge the granting of
Director's Discretionary Time for obtaining the NTT data. We made use of the
SIMBAD database, operated at CDS, Strasbourg, France. We thank the referee
Dr.\ Anders Winnberg for critical comments that helped improve the
presentation. Jacco thanks Joana Oliveira for help in reading funny formats of
data, improving this manuscript, and much more.
\end{acknowledgements}

\appendix

\section{H$_2$O maser emission from the Galactic AGB star R Doradus}

%
%
\begin{figure}[]
\centerline{\psfig{figure=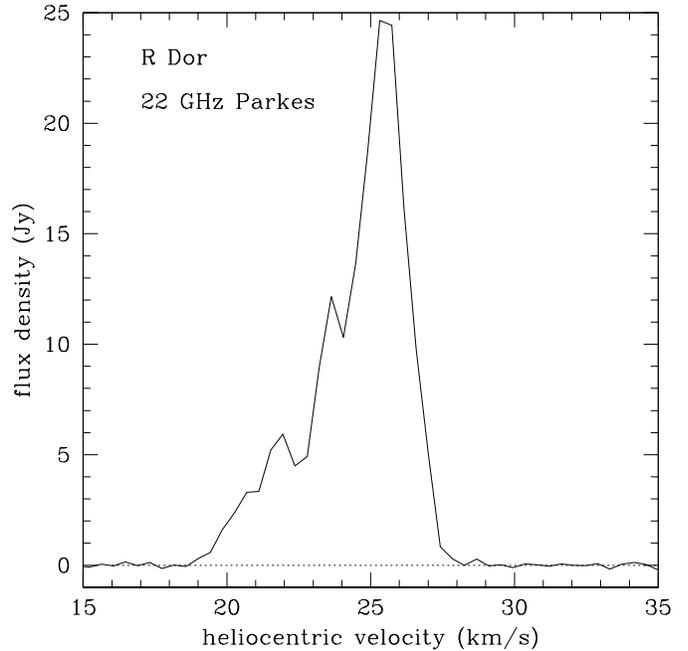,width=88mm}}
\caption[]{R Doradus: Circumstellar 22 GHz (Parkes 2000) H$_2$O maser
emission. The velocities are heliocentric.}
\end{figure}

R Doradus (IRAS04361$-$6210: $F_{12}=5157$, $F_{25}=1594$ Jy) is a famous
$6^{\rm th}$ magnitude (V-band) SRb variable AGB star with a period of 338
days and a spectral type M8 IIIe. It is at a distance of only 61 pc, making it
the biggest star on the night sky (Bedding et al.\ 1997 and references
therein).

The 22 GHz spectrum with Parkes is presented in Fig.\ A1, with an on-source
integration time of 228 seconds resulting in an rms noise of 74 mJy. The
(heliocentric) stellar restframe velocity is at $v_\star=+25$ km s$^{-1}$. The
blue-asymmetric CO($1\rightarrow0$) line profile indicates an expansion
velocity of $v_{\rm exp}=6.2$ km s$^{-1}$ (Lindqvist et al.\ 1992b) and a
mass-loss rate of $\dot{M}\sim10^{-7}$ M$_\odot$ yr$^{-1}$ (Loup et al.\ 1993).
The H$_2$O maser emission appears red-asymmetric, but as the peak very nearly
coincides with the stellar velocity there is actually more emission at the
blue-shifted rather than at the red-shifted side. Possibly the red-shifted
part of the emission is being occulted by the star. The blue-shifted emission
extends to (blue-shifted) velocities of almost $\sim6$ km s$^{-1}$ with
respect to the stellar restframe. The H$_2$O masers thus seem to trace
material that includes matter that has already nearly reached the final
outflow velocity.

R Dor is a good example of an AGB star with moderate mass loss. If placed at
the distance of the LMC, it would have been barely detectable by ISO in the
mid-IR (several mJy), and three orders of magnitude too faint to be detectable
at 22 GHz (few $\times10^{-2}$ mJy).

\section{Echelle 0.6 to 0.9 $\mu$m spectra of IRAS04553$-$6825}

%
%
\begin{figure}[b]
\centerline{\psfig{figure=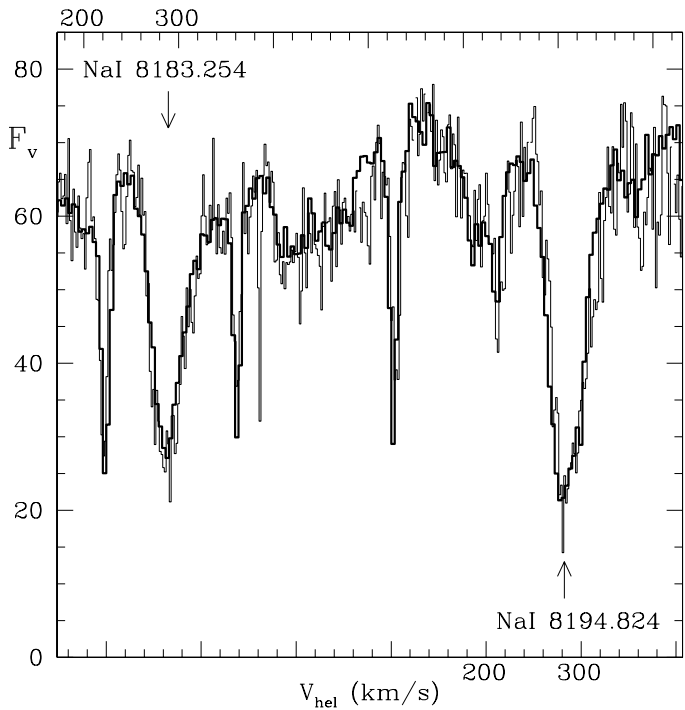,width=90mm}}
\caption[]{NTT echelle spectra of IRAS04553$-$6825 (October 1995, plus January
1996 in boldface) around Na {\sc i} $\lambda\lambda$8183,8195, with
corresponding heliocentric velocity axes. Flux units are arbitrary. The narrow
absorption lines are of telluric origin.}
\end{figure}

The 3.5 m New Technology Telescope (NTT) at the European Southern Observatory
(ESO) at La Silla, Chile, was used on October 7, 1995, and January 7, 1996,
with the ESO Multi-Mode Instrument (EMMI) to obtain echelle spectra of
IRAS04553$-$6825. Grating \#14 was used with grism \#6 as cross disperser in
October, and grism \#4 in January, yielding a spectral coverage from 6100 to
8300 \AA\ and from 6000 to 9000 \AA, respectively. The slit width and length
were $1^{\prime\prime}$ and $10^{\prime\prime}$ in October, and
$2^{\prime\prime}$ and $4^{\prime\prime}$ in January. The total integration
time was 1.5 hr in October, and 1 hr in January. The data were reduced in the
normal way using the Munich Interactive Data Analysis Software (MIDAS)
package. The wavelength calibration was done by taking a ThAr lamp spectrum in
conditions identical to the spectrum of IRAS04553$-$6825. The measured
spectral resolving power is $\sim7.5\times10^4$ for the spectrum taken in
October, and $\sim4\times10^4$ for the January spectrum. The average seeing
was $\sim1^{\prime\prime}$ on both nights.

Our echelle spectra resemble the lower resolution spectrum published by Elias
et al.\ (1986), confirming the spectral type of M7.5. Photometry obtained
using the 0.9 m Dutch telescope at ESO, La Silla, yield an approximate average
I-band magnitude $m_{\rm I}\sim13$ and $(V-I)\sim6$ mag. Van Loon et al.\
(1998b) concluded on the basis of the equivalent widths of the Ca {\sc ii}
triplet lines that IRAS04553$-$6825 has a typical LMC metallicity ($Z=0.008$).

%
%
\begin{figure}[b]
\centerline{\psfig{figure=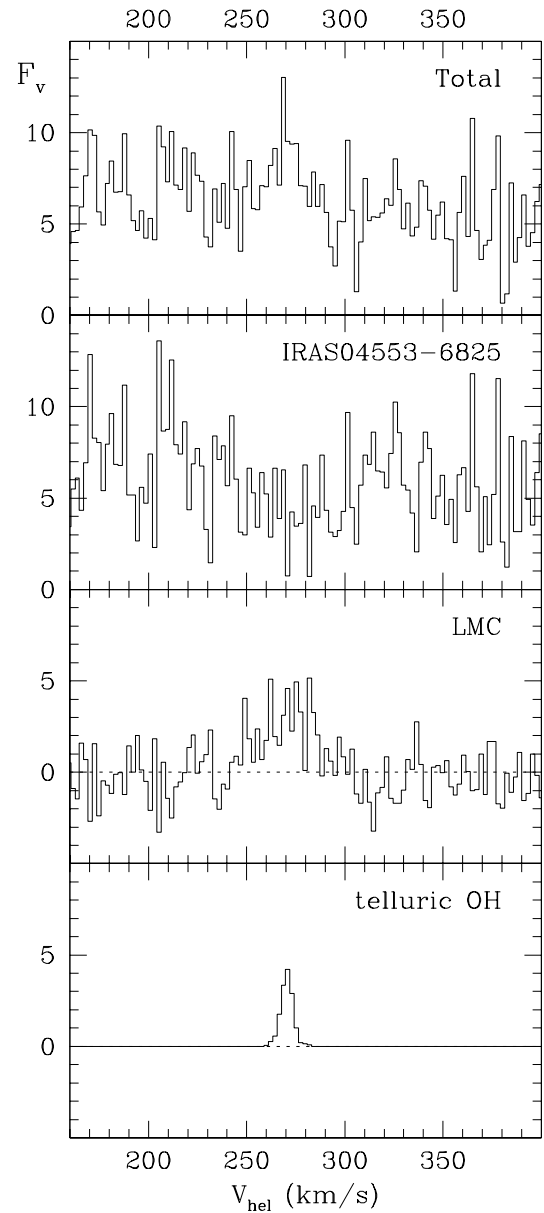,width=70mm}}
\caption[]{NTT echelle spectrum of IRAS04553$-$6825 around H$\alpha$,
deconvolved into the contributions of the photosphere of IRAS04553$-$6825, the
ISM of the LMC, and the telluric OH. Flux units are arbitrary. The
heliocentric velocities are for H$\alpha$.}
\end{figure}

The echelle spectrum of IRAS04553$-$6825 around the Na {\sc i}
$\lambda\lambda$8183,8195 lines is presented in Fig.\ B1, on an arbitrary flux
scale. Note the perfect match between the spectra taken with a time interval
of 3 months or 0.1 in pulsation phase, both between minimum and average light
in the K-band. The absorption is maximum at a heliocentric velocity between
280 and 290 km s$^{-1}$, whereas the K {\sc i} $\lambda$7699 and Ca {\sc ii}
$\lambda\lambda\lambda$8498,8542,8662 absorption is maximum at 295 km
s$^{-1}$. This is due to scattering of the photospheric spectrum by an
extended, expanding dust shell (see Romanik \& Leung 1981, and references
therein). The equivalent widths are measured as $W_{8183}=0.72\pm0.04$ \AA\
and $W_{8195}=0.88\pm0.04$ \AA\ in October 1995, and $W_{8183}=0.66\pm0.03$
\AA\ and $W_{8195}=0.82\pm0.02$ \AA\ in January 1996. Their sum is on average
$W_{8183+8195}=1.51\pm0.04$ \AA. The equivalent widths of the Na {\sc i} lines
reach their minima in giants, and are larger in both dwarfs and supergiants.
Comparison with Schiavon et al.\ (1997), and excluding a dwarf status,
confirms that IRAS04553$-$6825 has a supergiant-type optical spectrum.

Elias et al.\ (1986) claim the association with H$\alpha$ and forbidden line
emission from N$^+$ and S$^+$ ions. However, the red-shift of IRAS04553$-$6825
is such that the expected observed wavelengths of both the H$\alpha$, N {\sc
ii} $\lambda$6548, and S {\sc ii} $\lambda\lambda$6717,6731 emission lines
coincide within $\sim20$ km s$^{-1}$ with telluric OH emission lines. A
stretch of $\sim10$ \AA\ of photospheric spectrum around 6590 \AA\ that is
relatively free of molecular absorption is easily mistaken for the N {\sc ii}
$\lambda$6583 line in a lower resolution spectrum like that of Elias et al.

Other telluric OH lines in the spectrum were used to derive an empirical
relation between the intensity and quantum numbers associated with the
vibrational and rotational levels of the transition, making use of Osterbrock
\& Martel (1992) for identifications. A template line shape of the telluric OH
line was created by averaging 20 bright isolated lines. The telluric line was
then subtracted from the spectrum of the star (Fig.\ B2). What remains is
spatially extended H$\alpha$ emission covering $\sim50$ km s$^{-1}$, and weak
photospheric H$\alpha$ absorption. A 10 min exposure taken through an
H$\alpha$ filter with the NTT of the region around IRAS04553$-$6825
(http://www.eso.org/outreach/press-rel/pr-1996/phot-15-96.gif) shows extended
nebular complexes a few arcmin SW and faint filamentary emission due E of the
star, but no H$\alpha$ emission that can be attributed to the emission
detected in the spectrum at $\sim1$ pc from the star.

\section{Galactic Centre masers}

Wood et al.\ (1998), Blommaert et al.\ (1998) and Ortiz et al.\ (2000) have
searched for near- and mid-IR counterparts of OH sources in the direction of
the Galactic Centre, mainly from Lindqvist et al.\ (1992a). Their studies are
suitable for comparison of the properties of maser emission, expansion
velocities, pulsation periods and amplitudes, and CSE optical depths. A
distance to the Galactic Centre of 8 kpc is adopted (Reid 1993).

%
%
\begin{table*}
\caption[]{Identification of Galactic Centre AGB stars in the IRAS Point
Source Catalogue. Identifiers (L)WHM are from Lindqvist et al.\ (1992a) and
Wood et al.\ (1998), where the prefix S refers to single-peaked OH masers. The
PSC entry of a spatially coincident IRAS source is given, together with the
differences in Right Ascension ($\alpha$) and Declination ($\delta$), and the
total distance $\Delta$. The error ellipse of the IRAS detection is given by
major and minor axes $\sigma_{\rm a}$ and $\sigma_{\rm b}$ and angle $\theta$
indicating the orientation of the major axis, from North over East (i.e.\ in
these cases the major axis is roughly parallel to $\alpha$). Identification
(id) is decided positive (+) in case of agreement within the error ellipse,
and questionable (?) if not more than $3\sigma$ distance. Where available, we
quote from Lindqvist et al.\ (1992a) and Wood et al.\ (1998) the interstellar
extinction corrected $(H-K)$ and $K$ photometry, IRAS-PSC 12, 25, and 60
$\mu$m flux densities, bolometric magnitude $M_{\rm bol}$, pulsation period
$P$, and OH derived expansion velocity $v_\infty$ of the CSE. Underlined names
are sources with ISOGAL counterparts found by Ortiz et al.\ (2000).}
\begin{tabular}{ccrrrrrrccrrrrrrc}
\hline\hline
\llap{(}L)WHM                     &
IRAS-PSC                          &
$\Delta_{\alpha}$                 &
$\Delta_{\delta}$                 &
$\Delta$                          &
$\sigma_{\rm a}$                  &
$\sigma_{\rm b}$                  &
$\theta$                          &
id                                &
\llap{$($}$H-K$\rlap{$)$}         &
$K$                               &
$S_{12}$                          &
$S_{25}$                          &
$S_{60}$                          &
$M_{\rm bol}$                     &
$P$                               &
$v_\infty$                        \\
\hline
\llap{5}.                         &
17421$-$2931                      &
3                                 &
2                                 &
3                                 &
26                                &
5                                 &
93                                &
+                                 &
                                  &
                                  &
3\rlap{.3}                        &
5\rlap{.2}                        &
$<$30                             &
                                  &
                                  &
\llap{17}.\rlap{6}                \\
\llap{13}.                        &
17423$-$2924                      &
9                                 &
8                                 &
12                                &
34                                &
6                                 &
93                                &
?                                 &
                                  &
                                  &
4\rlap{.7:}                       &
16                                &
$<$362                            &
                                  &
                                  &
\llap{14}.\rlap{2}                \\
\llap{13}.\rlap{015}              &
17423$-$2924                      &
33                                &
16                                &
36                                &
34                                &
6                                 &
93                                &
?                                 &
1.29                              &
\llap{9}.43                       &
4\rlap{.7:}                       &
16                                &
$<$362                            &
$-2.33$                           &
213                               &
                                  \\
\llap{\underline{21}}.\rlap{031}  &
\underline{17413$-$2909}          &
1                                 &
1                                 &
2                                 &
31                                &
7                                 &
93                                &
+                                 &
3.02                              &
\llap{9}.27                       &
6\rlap{.5}                        &
$<$52                             &
$<$606                            &
$-5.26$                           &
698                               &
\llap{18}.\rlap{8}                \\
\llap{\underline{33}}.            &
\underline{17413$-$2903}          &
29                                &
2                                 &
29                                &
28                                &
6                                 &
93                                &
+                                 &
                                  &
                                  &
9\rlap{.0}                        &
$<$16                             &
$<$42                             &
                                  &
                                  &
\llap{15}.                        \\
\llap{\underline{34}}.\rlap{004}  &
\underline{17419$-$2907}          &
37                                &
27                                &
46                                &
26                                &
13                                &
93                                &
?                                 &
0.88                              &
\llap{7}.14                       &
4\rlap{.6}                        &
117                               &
467\rlap{:}                       &
$-4.69$                           &
453                               &
\llap{14}.\rlap{2}                \\
\llap{\underline{34}}.\rlap{063}  &
\underline{17419$-$2907}          &
55                                &
9                                 &
56                                &
26                                &
13                                &
93                                &
?                                 &
2.52                              &
\llap{9}.60                       &
4\rlap{.6}                        &
117                               &
467\rlap{:}                       &
$-3.70$                           &
915                               &
                                  \\
\llap{39}.\rlap{001}              &
17433$-$2918                      &
49                                &
2                                 &
49                                &
31                                &
6                                 &
93                                &
?                                 &
0.82                              &
\llap{7}.12                       &
$<$7\rlap{.3}                     &
6\rlap{.3}                        &
$<$77                             &
$-4.25$                           &
355                               &
                                  \\
\llap{39}.\rlap{002}              &
17433$-$2918                      &
50                                &
5                                 &
50                                &
31                                &
6                                 &
93                                &
?                                 &
2.23                              &
\llap{7}.53                       &
$<$7\rlap{.3}                     &
6\rlap{.3}                        &
$<$77                             &
$-4.69$                           &
559                               &
\llap{13}.\rlap{3}                \\
\llap{60}.\rlap{001}              &
17421$-$2857                      &
4                                 &
13                                &
14                                &
19                                &
5                                 &
93                                &
?                                 &
1.02                              &
\llap{6}.13                       &
16                                &
298                               &
$<$4318                           &
$-5.25$                           &
576                               &
                                  \\
\llap{\underline{64}}.\rlap{028}  &
\underline{17412$-$2849}          &
7                                 &
2                                 &
7                                 &
22                                &
6                                 &
93                                &
+                                 &
2.25                              &
\llap{8}.42                       &
4\rlap{.1}                        &
5\rlap{.8}                        &
$<$56                             &
$-5.70$                           &
692                               &
\llap{17}.\rlap{0}                \\
\llap{65}.\rlap{119}              &
17424$-$2859                      &
4                                 &
6                                 &
7                                 &
22                                &
7                                 &
93                                &
+                                 &
1.86                              &
\llap{8}.07                       &
1360                              &
5147                              &
18560                             &
$-5.01$                           &
799                               &
\llap{19}.\rlap{9}                \\
\llap{71}.\rlap{002}              &
17417$-$2851                      &
29                                &
7                                 &
30                                &
31                                &
9                                 &
93                                &
+                                 &
2.82                              &
\llap{9}.38                       &
11\rlap{:}                        &
54                                &
671\rlap{:}                       &
$-4.76$                           &
636                               &
\llap{24}.\rlap{4}                \\
\llap{\underline{79}}.\rlap{001}  &
\underline{17411$-$2843}          &
30                                &
3                                 &
30                                &
32                                &
5                                 &
\llap{1}02                        &
+                                 &
1.07                              &
\llap{6}.48                       &
3\rlap{.4}                        &
4\rlap{.2}                        &
$<$53                             &
$-5.02$                           &
477                               &
\llap{9}.\rlap{7}                 \\
\llap{82}.\rlap{007}              &
17428$-$2854                      &
16                                &
2                                 &
16                                &
23                                &
7                                 &
93                                &
+                                 &
1.83                              &
\llap{7}.99                       &
12                                &
151\rlap{:}                       &
$<$57                             &
$-4.89$                           &
701                               &
\llap{23}.\rlap{9}                \\
\llap{\underline{101}}.\rlap{014} &
\underline{17419$-$2837}          &
61                                &
1                                 &
61                                &
74                                &
7                                 &
93                                &
+                                 &
3.41                              &
\llap{10}.33                      &
$<$10                             &
4\rlap{.3}                        &
$<$99                             &
$-4.98$                           &
825                               &
\llap{15}.\rlap{8}                \\
\llap{134}.                       &
17436$-$2807                      &
1                                 &
8                                 &
8                                 &
47                                &
5                                 &
93                                &
?                                 &
                                  &
                                  &
4\rlap{.7}                        &
$<$30                             &
$<$451                            &
                                  &
                                  &
\llap{16}.\rlap{3}                \\
\llap{\underline{S02}}.           &
17428$-$2918                      &
39                                &
1                                 &
39                                &
46                                &
10                                &
92                                &
+                                 &
                                  &
                                  &
2\rlap{.8:}                       &
6\rlap{.3}                        &
$<$212                            &
                                  &
                                  &
                                  \\
\llap{\underline{S07}}.           &
17429$-$2903                      &
55                                &
17                                &
58                                &
30                                &
7                                 &
\llap{1}02                        &
?                                 &
                                  &
                                  &
$<$7\rlap{.6}                     &
23                                &
548\rlap{:}                       &
                                  &
                                  &
                                  \\
\llap{\underline{S15}}.           &
17441$-$2822                      &
7                                 &
3                                 &
7                                 &
23                                &
6                                 &
93                                &
+                                 &
                                  &
                                  &
$<$7\rlap{.7}                     &
69                                &
12952                             &
                                  &
                                  &
                                  \\
\hline
\end{tabular}
\end{table*}

Cross-identifications of the samples of stars in Lindqvist et al.\ (1992a) and
Wood et al.\ (1998) with the IRAS Point Source Catalogue (PSC) are summarised
in Table C1. Where the OH source and/or LPV lies inside of the error ellipse
as given in the PSC, the identification was tagged positive (+). If the
distances along both axes of the error ellipse are less than $3\sigma$, the
identification was tagged questionable (?). This resulted in 11 likely IRAS
counterparts. All of these that have known near-IR counterparts have IR
colours that suggest oxygen-rich CSEs (from $(K-[12])$ versus $(H-K)$, see van
Loon et al.\ 1998a). The identification of 65.119 with IRAS17424$-$2859 is
uncertain because the 12 $\mu$m emission is much too bright to arise from a
CSE around an evolved star given its near-IR colours, and this object is
further ignored. Ortiz et al.\ (2000) searched for 7 and 15 $\mu$m
counterparts of OH/IR stars in the Galactic Centre using the ISOGAL (Omont et
al.\ 1999) database. All of their identifications of Lindqvist sources with
IRAS sources are also recovered by us, but there are at least as many more
identifications that they did not find. The ISOGAL survey had to avoid strong
IRAS point sources and is therefore biased, whereas the IRAS survey and our
approach of identifying OH/IR stars are expected to be more homogeneous.

The IRAS 12 $\mu$m flux densities of the positively identified sources are all
in the range 2.8 to 12 Jy. This corresponds to flux densities from 0.07 to 0.3
Jy at the distance of the LMC. OH/IR stars with IRAS 12 $\mu$m flux densities
corresponding to $\sim1$ Jy at the LMC are not encountered in the Galactic
Centre, but do exist in the MCs (e.g.\ Table 2). This is especially surprising
as it has been suggested that among the Galactic Centre OH/IR stars are
massive, metal-rich mass-losing LPVs that are expected to have a large IR
excess emission from their dusty CSEs. At the other extreme, if most of the
Galactic Centre CSEs have 12 $\mu$m flux densities lower than the positively
identified IRAS sources, they would have been largely undetected by IRAS in
the MCs.

Wood et al.\ (1998) find LPVs with K-band magnitudes from 5 to 13 after
correction for interstellar extinction, with the faintest IRAS associated LPV
to have $K_0=10.33$ mag. At the distance of the LMC this would yield K-band
magnitudes from 9 to 17, so most of these would also have been found by the
IRAC2 searches for near-IR counterparts of IRAS sources in the MCs by Zijlstra
et al.\ 1996, van Loon et al.\ 1997, and Groenewegen \& Blommaert (1998). The
$(H-K)_0$ colours of the Galactic Centre LPVs are $\sim2$ mag on average and
$\sim4$ mag maximum, similar to the $(H-K)$ colours of the obscured AGB stars
in the MCs.

%
%
\begin{figure}[b]
\centerline{\psfig{figure=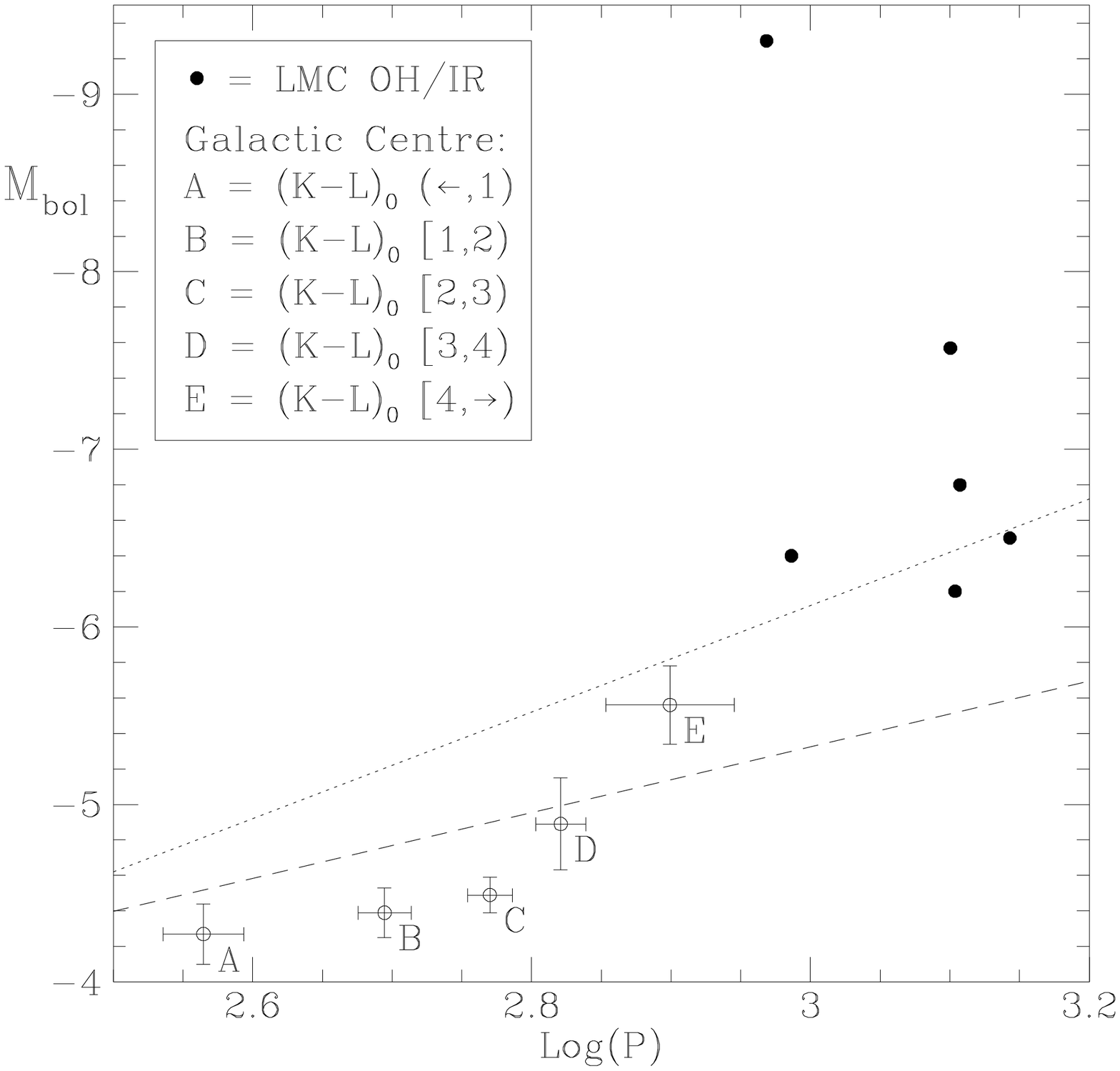,width=88mm}}
\caption[]{Period-luminosity diagram for OH/IR stars in the LMC and Galactic
Centre --- the latter grouped according to near-IR colour. The lines are
relations derived from oxygen (dotted) and carbon-rich (dashed) LPVs with
$P<420$ d.}
\end{figure}

The mean values of $\log(P)$ and $M_{\rm bol}$ according to the $(K-L)_0$
colour (corrected for interstellar extinction) for the LPVs in the Galactic
Centre follow a sequence in the period-luminosity (P-L) diagram (Fig.\ C1).
Also plotted are the OH maser sources in the LMC that have known pulsation
periods and bolometric magnitudes, and the P-L relations that have been
derived from samples of oxygen-rich LPVs (dotted) and carbon-rich LPVs
(dashed) with periods $P<420$ d (Feast et al.\ 1989). The LMC OH maser stars
with known pulsation period have typically $(K-L)\sim2$ mag. Yet they appear
to be on the extension of the $(K-L)$ colour sequence as traced by the
Galactic Centre LPVs, despite that this sequence already reaches $(K-L)>4$ mag
before reaching the long periods and high luminosities of the LMC stars. This
illustrates that LMC stars with a similar stellar structure as Galactic Centre
stars have less dusty CSEs due to their lower metallicity (see also van Loon
2000).


\begin{thebibliography}{}
\bibitem[1990]{}
Alcolea J., Bujarrabal V., G\'{o}mez-Gonz\'{a}lez J., 1990, A\&A 231, 431
\bibitem[1999]{}
Alcolea J., Pardo J.R., Bujarrabal V., et al., 1999, A\&AS 139, 461
\bibitem[1997]{}
Bedding T.R., Zijlstra A.A., van der L\"{u}he O., et al., 1997, MNRAS 286, 957
\bibitem[1999]{}
Bica E.L.D., Schmitt H.R., Dutra C.M., Oliveira H.L., 1999, AJ 117, 238
\bibitem[1998]{}
Blommaert J.A.D.L., van der Veen W.E.C.J., van Langevelde H.J., Habing H.J.,
Sjouwerman L.O., 1998, A\&A 329, 991
\bibitem[1974]{}
Bohannan B., Epps H.W., 1974, A\&AS 18, 47
\bibitem[1996]{}
Bujarrabal V., Alcolea J., S\'{a}nchez-Contreras C., Colomer F., 1996, A\&A
314, 883
\bibitem[1998]{}
Castilho B.V., Gregorio-Hetem J., Spite F., Spite M., Barbuy B., 1998, A\&AS
127, 139
\bibitem[1997]{}
Cernicharo J., Alcolea J., Baudry A., Gonz\'{a}lez-Alfonso E., 1997, A\&A 319,
607
\bibitem[2000]{}
Cioni M.-R., Loup C., Habing H.J., 2000, A\&AS 144, 235
\bibitem[2000]{}
Colomer F., Reid M.J., Menten K.M., Bujarrabal V., 2000, A\&A 355, 979
\bibitem[1984]{}
Diamond P.J., Norris R.P., Booth R.S., 1984, MNRAS 207, 611
\bibitem[1986]{}
Elias J.H., Frogel J.A., Schwering P.B.W., 1986, ApJ 302, 675
\bibitem[1989]{}
Feast M.W., Glass I.S., Whitelock P.A., Catchpole R.M., 1989, MNRAS 241, 375
\bibitem[1982]{}
Fehrenbach C., Duflot M., 1982, A\&AS 48, 409
\bibitem[1976]{}
Goldreich P., Scoville N.Z., 1976, ApJ 205, 144
\bibitem[1998]{}
Gonz\'{a}lez-Alfonso E., Cernicharo J., Alcolea J., Orlandi M.A., 1998, A\&A
334, 1016
\bibitem[1998]{}
Groenewegen M.A.T., Blommaert J.A.D.L., 1998, A\&A 332, 25
\bibitem[1995]{}
Groenewegen M.A.T., Smith C.H., Wood P.R., Omont A., Fujiyoshi T., 1995, ApJ
449, L119
\bibitem[1998]{}
Groenewegen M.A.T., Whitelock P.A., Smith C.H., Kerschbaum F., 1998, MNRAS
293, 18
\bibitem[1990]{}
Haikala L.K., 1990, A\&AS 85, 875
\bibitem[1994]{}
Haikala L.K., Nyman L.\AA., Forsstr\"{o}m V., 1994, A\&AS 103, 107
\bibitem[1974]{}
Harvey P.M., Bechis K.P., Wilson W.J., Ball J.A., 1974, ApJS 27, 331
\bibitem[1956]{}
Henize K.G., 1956, ApJS 2, 315
\bibitem[1998]{}
Herpin F., Baudry A., Alcolea J., Cernicharo J., 1998, A\&A 334, 1037
\bibitem[1982]{}
Huggins P.J., Glassgold A.E., 1982, AJ 87, 1828
\bibitem[1994]{}
Izumiura H., Deguchi S., Hashimoto O., et al., 1994, ApJ 437, 419
\bibitem[1999]{}
Kim S., Dopita M.A., Staveley-Smith L., Bessell M.S., 1999, AJ 118, 2797
\bibitem[1997]{}
Kwok S., Volk K., Bidelman W.P., 1997, ApJS 112, 557
\bibitem[1992]{}
Laval A., Rosado M., Boulesteix J., et al., 1992, A\&A 253, 213
\bibitem[1994]{}
Laval A., Gry C., Rosado M., Marcelin M., Greve A., 1994, A\&A 288, 572
\bibitem[1993]{}
Le Bertre T., 1993, A\&AS 97, 729
\bibitem[1990]{}
Le Bertre T., Nyman L.-\AA., 1990, 233, 477
\bibitem[1989]{}
Lewis B.M., 1989, ApJ 338, 234
\bibitem[1991]{}
Lewis B.M., 1991, AJ 101, 254
\bibitem[1998]{}
Lewis B.M., 1998, ApJ 508, 831
\bibitem[1990]{}
Lindqvist M., Winnberg A., Forster J.R., 1990, A\&A 229, 165
\bibitem[1991]{}
Lindqvist M., Ukita N., Winnberg A., Johansson L.E.B., 1991, A\&A 250, 431
\bibitem[1992a]{}
Lindqvist M., Habing H.J., Winnberg A., 1992a, A\&A 259, 118
\bibitem[1992b]{}
Lindqvist M., Olofsson H., Winnberg A., Nyman L.-\AA, 1992b, A\&A 263, 183
\bibitem[1993]{}
Loup C., Forveille T., Omont A., Paul J.F., 1993, A\&AS 99, 291
\bibitem[1997]{}
Loup C., Zijlstra A.A., Waters L.B.F.M., Groenewegen M.A.T., 1997, A\&AS 125,
419
\bibitem[1992]{}
Luks Th., Rohlfs K., 1992, A\&A 263, 41
\bibitem[1980]{}
Meaburn J., 1980, MNRAS 192, 365
\bibitem[1993]{}
Meyssonnier N., Azzopardi M., 1993, A\&AS 102, 451
\bibitem[1983]{}
Morris M., Jura M., 1983, ApJ 267, 179
\bibitem[1993]{}
Netzer N., Elitzur M., 1993, ApJ 410, 701
\bibitem[1984]{}
Norris R.P., Booth R.S., Diamond P.J., et al., 1984, MNRAS 208, 435
\bibitem[1985]{}
Nyman L.-\AA., Olofsson H., 1985, A\&A 147, 309
\bibitem[1986]{}
Nyman L.-\AA., Olofsson H., 1986, A\&A 158, 67
\bibitem[1993]{}
Nyman L.-\AA, Hall P.J., Le Bertre T., 1993, A\&A 280, 551
\bibitem[1998]{}
Nyman L.-\AA, Hall P.J., Olofsson H., 1998, A\&AS 127, 185
\bibitem[1999]{}
Omont A., The ISOGAL Collaboration, 1999, in: The Universe as Seen by ISO,
eds.\ P. Cox \& M.F. Kessler. ESA-SP 427, p211
\bibitem[2000]{}
Ortiz R., Blommaert J.A.D.L., Copet E., et al., 2000, submitted to A\&A Main
Journal
\bibitem[1992]{}
Osterbrock D.E., Martel A., 1992, PASP 104, 76
\bibitem[1992]{}
Parker J.W., Garmany C.D., Massey P., Walborn N.R., 1992, AJ 103, 1205
\bibitem[1994]{}
Persi P., Palagi F., Felli M., 1994, A\&A 291, 577
\bibitem[1989]{}
Pr\'{e}vot L., Rousseau J., Martin N., 1989, A\&A 225, 303
\bibitem[1993]{}
Rebeirot E., Azzopardi M., Westerlund B.E., 1993, A\&AS 97, 603
\bibitem[1993]{}
Reid M.J., 1993, ARA\&A 31, 345
\bibitem[1990]{}
Reid I.N., Tinney C.G., Mould J.R., 1990, ApJ 348, 98
\bibitem[1996]{}
Richards A.M.S., Yates J.A., Cohen R.J., 1996, MNRAS 282, 665
\bibitem[1981]{}
Romanik C.J., Leung C.M., 1981, ApJ 246, 935
\bibitem[1996]{}
Rosado M., Laval A., Le Coarer E., et al., 1996, A\&A 308, 588
\bibitem[1997]{}
Schiavon R.P., Barbuy B., Rossi S.C.F., Milone A., 1997, ApJ 479, 902
\bibitem[1999]{}
Sivagnanam P., David P., 1999, MNRAS 304, 622
\bibitem[1989]{}
Sivagnanam P., Le Squeren A.M., Foy F., Tran Minh F., 1989, A\&A 211, 341
\bibitem[1990]{}
Sivagnanam P., Le Squeren A.M., Biraud F., Diamond P.J., 1990, A\&A 229, 171
\bibitem[1998]{}
Sjouwerman L.O., van Langevelde H.J., Winnberg A., Habing H.J., 1998, A\&AS
128, 35
\bibitem[1995]{}
Smith V.V., Plez B., Lambert D.L., 1995, ApJ 441, 735
\bibitem[1999]{}
Stanimirovi\'{c} S., Staveley-Smith L., Dickey J.M., Sault R.J., Snowden S.L.,
1999, MNRAS 302, 417
\bibitem[1997]{}
Staveley-Smith L., Sault R.J., Hatzidimitriou D., Kesteven M., McConnell D.,
1997, MNRAS 289, 225
\bibitem[1994]{}
Takaba H., Ukita N., Miyaji T., Miyoshi M., 1994, PASJ 46, 629
\bibitem[1999]{}
Trams N.R., van Loon J.Th., Waters L.B.F.M., et al., 1999, A\&A 346, 843
\bibitem[2000]{}
van Loon J.Th., 2000, A\&A 354, 125
\bibitem[2000]{}
van Loon J.Th., Zijlstra A.A., 2000, ApJL in press
\bibitem[1996]{}
van Loon J.Th., Zijlstra A.A., Bujarrabal V., Nyman L.-{\AA}., 1996, A\&A 306,
L29
\bibitem[1997]{}
van Loon J.Th., Zijlstra A.A., Whitelock P.A., et al., 1997, A\&A 325, 585
\bibitem[1998a]{}
van Loon J.Th., Zijlstra A.A., Whitelock P.A., et al., 1998a, A\&A 329, 169
\bibitem[1998b]{}
van Loon J.Th., te Lintel Hekkert P., Bujarrabal V., Zijlstra A.A., Nyman
L.-\AA, 1998b, A\&A 337, 141
\bibitem[1999a]{}
van Loon J.Th., Zijlstra A.A., Groenewegen M.A.T., 1999a, A\&A 346, 805
\bibitem[1999b]{}
van Loon J.Th., Groenewegen M.A.T., de Koter A., et al., 1999b, A\&A 351, 559
\bibitem[2000]{}
van Loon J.Th., Zijlstra A.A., Kaper L., et al., 2000, A\&A in press
\bibitem[1981]{}
Westerlund B.E., Olander N., Hedin B., 1981, A\&AS 43, 267
\bibitem[1989]{}
Whitelock P.A., Feast M.W., Menzies J.W., Catchpole R.M., 1989, MNRAS 238, 769
\bibitem[1986]{}
Whiteoak J.B., Gardner F.F., 1986, MNRAS 222, 513
\bibitem[1998]{}
Wood P.R., 1998, A\&A 338, 592
\bibitem[1983]{}
Wood P.R., Bessell M.S., Fox M.W., 1983, ApJ 272, 99
\bibitem[1986]{}
Wood P.R., Bessell M.S., Whiteoak J.B., 1986, ApJ 306, L81
\bibitem[1992]{}
Wood P.R., Whiteoak J.B., Hughes S.M.G., et al., 1992, ApJ 397, 552
\bibitem[1998]{}
Wood P.R., Habing H.J., McGregor P.J., 1998, A\&A 336, 925
\bibitem[1997]{}
Yates J.A., Field D., Gray M.D., 1997, MNRAS 285, 303
\bibitem[1996]{}
Zijlstra A.A., Loup C., Waters L.B.F.M., et al., 1996, MNRAS 279, 32
\bibitem[2000]{}
Zijlstra A.A., Chapman J.M., te Lintel Hekkert P., et al., 2000, MNRAS in
press
\end{thebibliography}
\end{document}